\documentclass[english]{article}
\usepackage[T1]{fontenc}
\usepackage[latin1]{inputenc}
\usepackage{float}
\usepackage{calc}
\usepackage{graphicx}
\usepackage{amssymb}

\makeatletter

\providecommand{\tabularnewline}{\\}

\usepackage{float}

\usepackage{calc}

\makeatletter

\usepackage{float}

\usepackage{calc}

\makeatletter

\usepackage{float}

\usepackage{calc}

\usepackage{color}

\makeatletter

\usepackage{float}

\usepackage{calc}



\makeatother

\makeatother

\makeatother

\usepackage{babel}
\makeatother
\begin{document}
\sloppy

\title{Stability of Localized Patterns in Neural Fields}

\author{Konstantin Doubrovinski$^{1}$%
\thanks{present address: Universität des Saarlandes, Postfach 151150, 66041
Saarbrücken%
}, J.~Michael Herrmann$^{2}$\\[4mm] $^{1}$Max Planck Institute
for the Physics of Complex Systems\\
 Div.~Biological Physics, Nöthnitzer Str.~38, 01187 Dresden, Germany\\
 $^{2}$Göttingen University, Institute for Nonlinear Dynamics\\
 and Bernstein Center for Computational Neuroscience\\
 Bunsenstr.~10, 37073 Göttingen, Germany \\[2mm] kodo7115@mpipks-dresden.mpg.de,~michael@nld.ds.mpg.de}

\maketitle

\begin{abstract}
We investigate two-dimensional neural fields as a model of the dynamics
of macroscopic activations in a cortex-like neural system. While the
one-dimensional case has been treated comprehensively by Amari 30
years ago, two-dimensional neural fields are much less understood.
We derive conditions for the stability for the main classes of localized
solutions of the neural field equation and study their behavior beyond
parameter-controlled destabilization. We show that a slight modification
of original model yields an equation whose stationary states are guaranteed
to satisfy the original problem and numerically demonstrate that it
admits localized non-circular solutions. Generically, however, only
periodic spatial tessellations emerge upon destabilization of rotationally-invariant
solutions. 
\end{abstract}

\section{Introduction}

Neural fields (Amari 1977) describe the dynamics of distributions
of activity on a layer of neurons. Neural fields have been suggested
as models of internal representations in natural agents (Takeuchi
and Amari 1979, Gross \emph{et al.}~1998) as well as in robots (Steinhage
2000, Iossifidis and Steinhage 2001, Erlhagen and Bicho 2006). Various
modalities are covered such as spatial localization, viewing direction,
attentional spotlight, the dynamics of decision making, elementary
behaviors, and positions of other agents in the environment. More
extensive studies in theoretical neuroscience (Suder \emph{et al.}~1999,
Mayer \emph{et al.}~2002, Bressloff 2006) concentrate on primary
visual cortex, cf.~(Lieke \emph{et al.}~1989), superior culliculus
(Schierwagen and Werner 1996), the representation of motoric primitives
(Thelen \emph{et al.}~2001, Erlhagen and Schöner 2002), and working
memory in prefrontal cortex (Schutte \emph{et al.}~2003, Camperi
and Wang 1998).

Recently neural fields are experiencing attention in modeling and
analysis of brain imaging data, because they are able to represent
the dynamic interaction of an active medium with time-varying inputs,
and because the spatial and temporal scales in the data and neural
field models \textcolor{black}{are starting to become comparable}.
Especially if information about connectivity is available from the
data (Jirsa \emph{et al.}~2002) then neural fields are of high explanatory
power.

Moreover the analysis has reached a level where applications directly
benefit from the theoretical progress, and at the same time, computational
power became available that allows us to perform on-line simulations
of two-dimensional neural fields.

Generally speaking, neural fields serve as nonparametric representations
of probability densities and their dynamics may perform operations
on the densities such as Bayesian computations (Herrmann \emph{et
al.}~1999). One-dimensional problems have been comprehensively studied
already in the 1970s (Amari 1977, Kishimoto and Amari 1979, Takeuchi
and Amari 1979). While for spatially extended, e.g.~periodic patterns,
the transition to the more relevant two-dimensional case is nontrivial
but fairly well understood (Ermentrout and Cowan 1979, Ermentrout
1998), localized activities in dimensions larger than one have not
yet been treated with the same rigor. The situation, however, does
not seem to be exactly complex: a large body of numerical studies
together with theoretical considerations (Laing \emph{et al.}~2002,
Laing and Troy 2003) imply a general instability of multi-bump solutions
if the interactions are excitatory at small and inhibitory at large
distances (see also Laing and Chow (2001) for stability analysis in
one-dimensional models of spiking neurons). Further, there has been
numerical evidence that single-bump solutions in two dimensions for
radially symmetric interactions are essentially circular, which was
exploited as an assumption in Taylor's early attack to the two-dimensional
case (Taylor 1999). In ref.~(Werner and Richter 2001) evidence has
been provided for the existence of ring-shaped solutions which are
possible for certain types of neural interactions. Along these lines
one may conjecture that finite mesh-like structures of higher genus
do exist as well.

The situation became more spirited only recently when in Refs.~(Herrmann
\emph{et al.}~2004, Bressloff 2005, Doubrovinski 2005) the stability
problem of localized activations in two-dimensional fields was eventually
tackled. Although the generality which has been achieved in the one-dimensional
neural fields is presently out of reach in two dimensions, a number
of interesting variants of the circular activity configurations were
analyzed so far.

Here we present a concise and reproducible scheme for the analysis
of the stability of localized activity distributions in neural fields.
We show the applicability of the scheme not only to the special case
of circular solution but study also ring-shaped and bar-shaped solutions
which present the complete set of known localized solutions for simple
kernels (Werner and Richter 2001). In addition to the stability proofs
which are based on the classical scheme (Pismen 1999, Bressloff 2005,
Doubrovinski 2005), we are interested mainly in the behavior beyond
the phase transitions towards the unstable regions. The destabilization
of circular solution is known to lead to a transient elongation of
the activity bubble (Bressloff 2005, Doubrovinski 2005) which ultimately
causes the localized solution to split or to form meandering bands.
Either case is unstable in a strict sense: the splitting into two
continues toward a plane-filling hexagonal pattern while the banded
patterns develop a global stripe pattern or a quasiperiodic arrangement.

The destabilization is thus fundamental since for typical Mexican-hat
interaction kernels there is no nearby stable state which is approached
after the bifurcation, while the spatially extended patterns are not
approached in finite time (unless a general criterion for convergence
is drawn into consideration). Yet the destabilization, at least in
the neurobiological applications is the most interesting part of the
theory. Divergences are usually very slow and may halt completely
due to reasonable boundary conditions (cf.~below), such that an activity-based
correlational learning scheme may organize anisotropies in the connections
which stabilizes the anisotropic activities as assumed in Ref.~(Bressloff
2005) and exploited in (Schierwagen and Werner 1996). A theoretical
account of the interaction of activity dynamics and learning was studied
in (Dong and Hopfield 1992), in relation to activity effect on feature
maps cf.~(Mayer \emph{et al.}~2002).

\section{The neural field equation}

The neural field model describes the activations of a layer of neurons
when the geometry of interactions rather than the specific connectivity
among the neurons is relevant. We assume positions~$\mathbf{r}\in\mathbb{R}^{2}$
for neurons with continuous-valued activations~$u\left(\mathbf{r},t\right)$.
The synaptic weights between neurons at the positions~$\mathbf{r}$
and~$\mathbf{r'}$ is expressed by isotropic interaction kernel~$w\left(\mathbf{r},\mathbf{r'}\right)=w\left(\left|\mathbf{r}-\mathbf{r'}\right|\right)$
of Mexican-hat shape. Neurons are activated if their total input is
greater than zero. We will study only equilibrium solutions without
external input, and we neglect slow learning effects, so the synaptic
weights are constant over time.

The activation at a position results from a weighted integration over
the inputs from all other active locations in the field and a natural
decay towards a resting potential denoted by $h$. The dynamics of
the neural field is thus determined by the equation \begin{equation}
\tau\frac{\partial u\left(\mathbf{r},t\right)}{\partial t}=-u\left(\mathbf{r},t\right)+\int_{R\left[u\right]}w\left(\left|\mathbf{r}-\mathbf{r'}\right|\right)d\mathbf{r'}+h,\label{eq: Amari eq}\end{equation}
 where~$R\left[u\right]=\left\{ x\mid u\left(\mathbf{r}\right)>0\right\} $
is the excited region, i.e.~a neuron receives input only from neurons
within $R$. The boundary of~$R\left[u\right]$ is assumed to be
smooth. Rescaling time,~$\tau$ can be set to unity without loss
of generality. Equilibrium solutions are defined by\begin{equation}
u\left(\mathbf{r},t\right)=\int_{R\left[u\right]}w\left(\left|\mathbf{r}-\mathbf{r'}\right|\right)d\mathbf{r'}+h\label{eq: Stat states}\end{equation}
 and depend on the value of the threshold parameter $h$ and the particular
form of $w$. Here we use a smooth kernel which is more general than
the quasi-constant kernel function in Ref.~(Herrmann \emph{et al.}~2004).
The kernel is constructed as a difference of Gaussian functions and
is defined by four parameters $K$, $k$, $M$, $m$: \begin{equation}
w=K\exp\left(-k\left\Vert \mathbf{r}-\mathbf{r'}\right\Vert ^{2}\right)-M\exp\left(-m\left\Vert \mathbf{r}-\mathbf{r'}\right\Vert ^{2}\right)\label{eq: Kernel def}\end{equation}

If non-rotationally symmetric solutions are excluded from the beginning
from the consideration of Eq.~\ref{eq: Stat states}, the situation
simplifies dramatically and it can be shown (Taylor 1999) that one-bump
solutions $u\left(\left\Vert \mathbf{r}\right\Vert \right)$ of certain
radii are stationary states of the dynamics (\ref{eq: Amari eq}).
Analogously, a ring-shaped solution (Werner and Richter 2001) or a
stripe-like solution, i.e.~a degenerate ring of infinite radius,
can occur. However, when considering an arbitrarily small perturbation
of the solution, the symmetry might be broken and new phenomena can
appear, as will be studied in the following.

\section{Stability}

\begin{figure}[H]
\begin{centering}\begin{tabular}{cc}
a\includegraphics[scale=0.33]{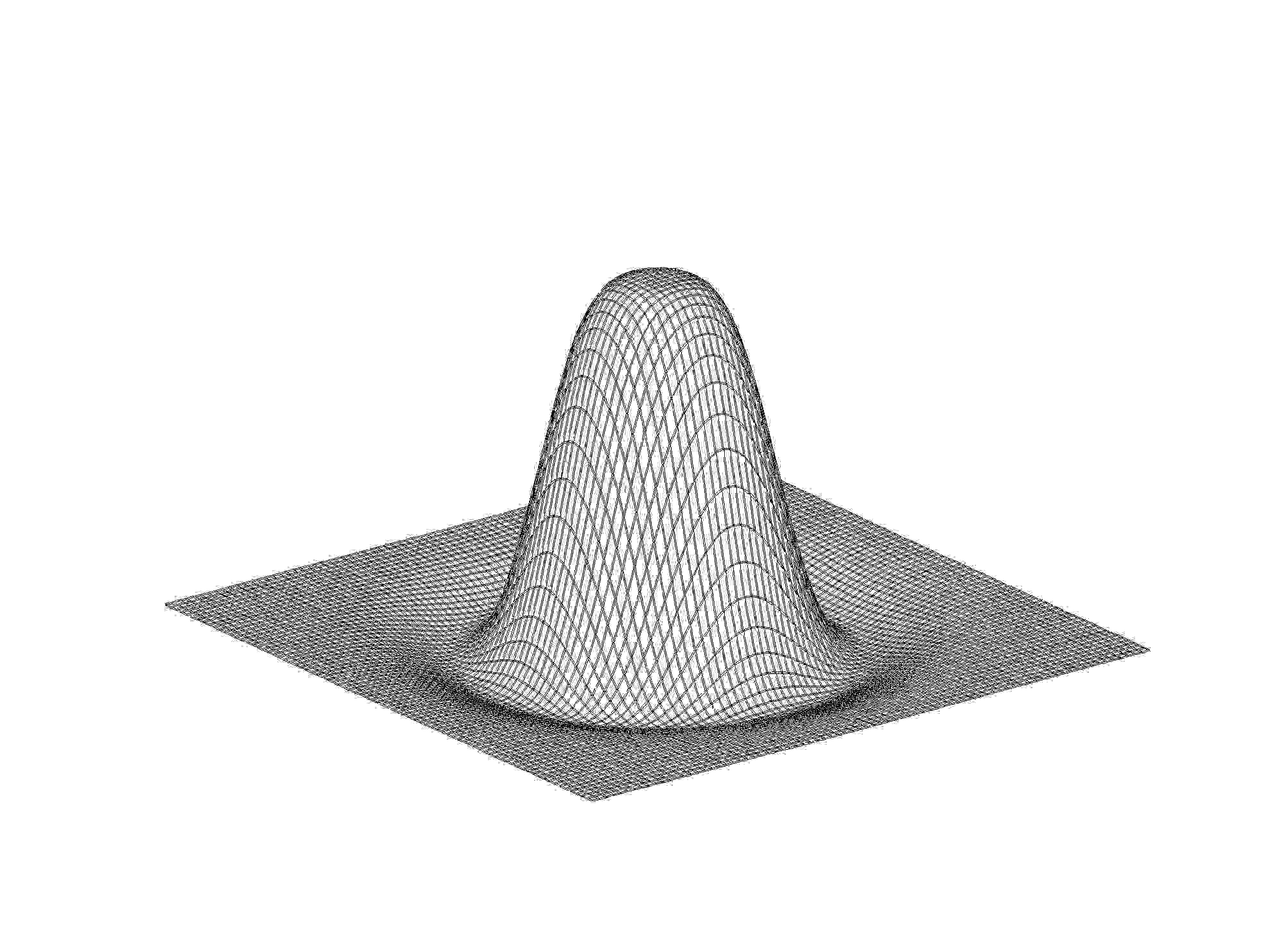}&
b\includegraphics[scale=0.33]{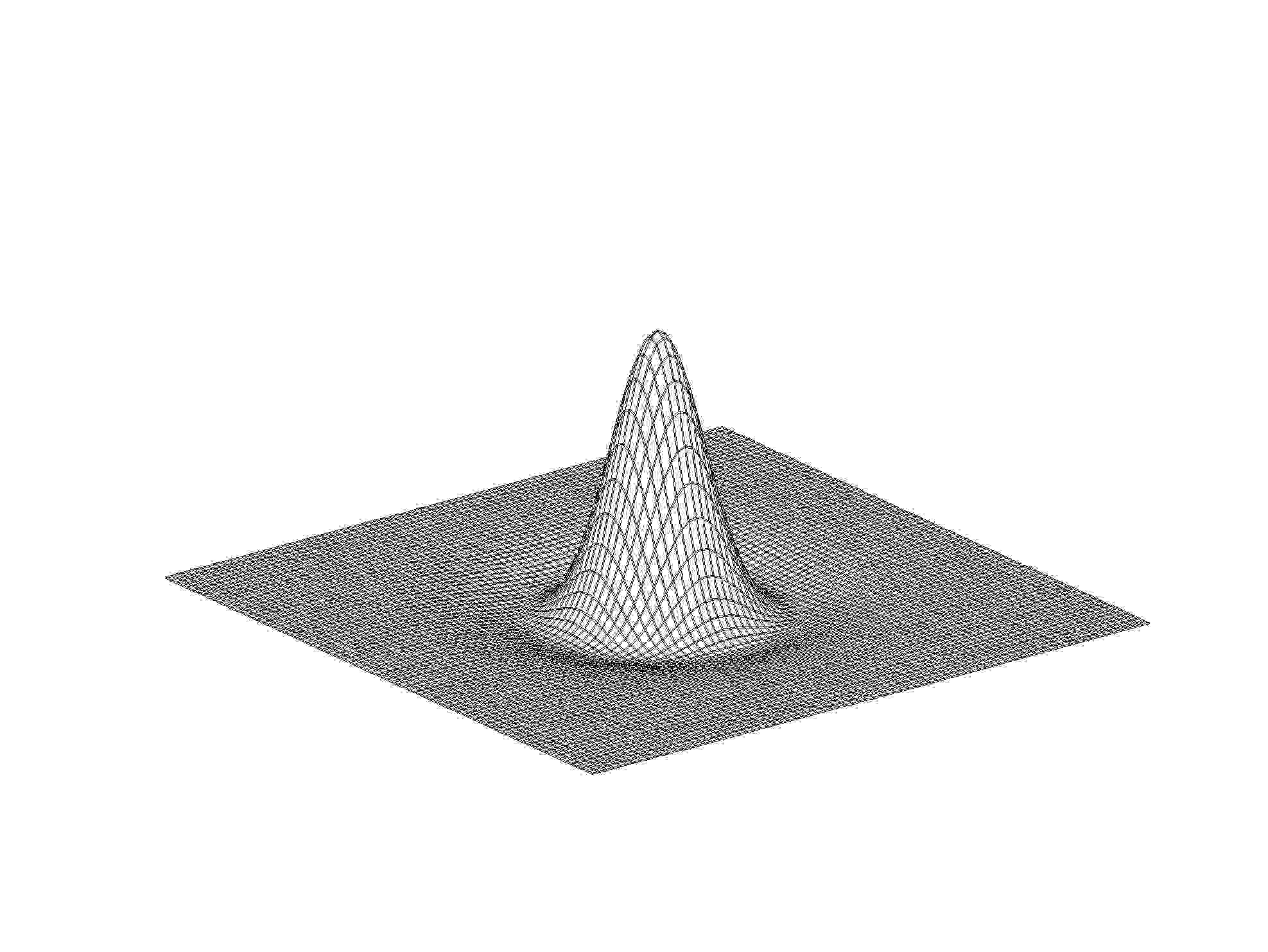}\tabularnewline
c\includegraphics[scale=0.2]{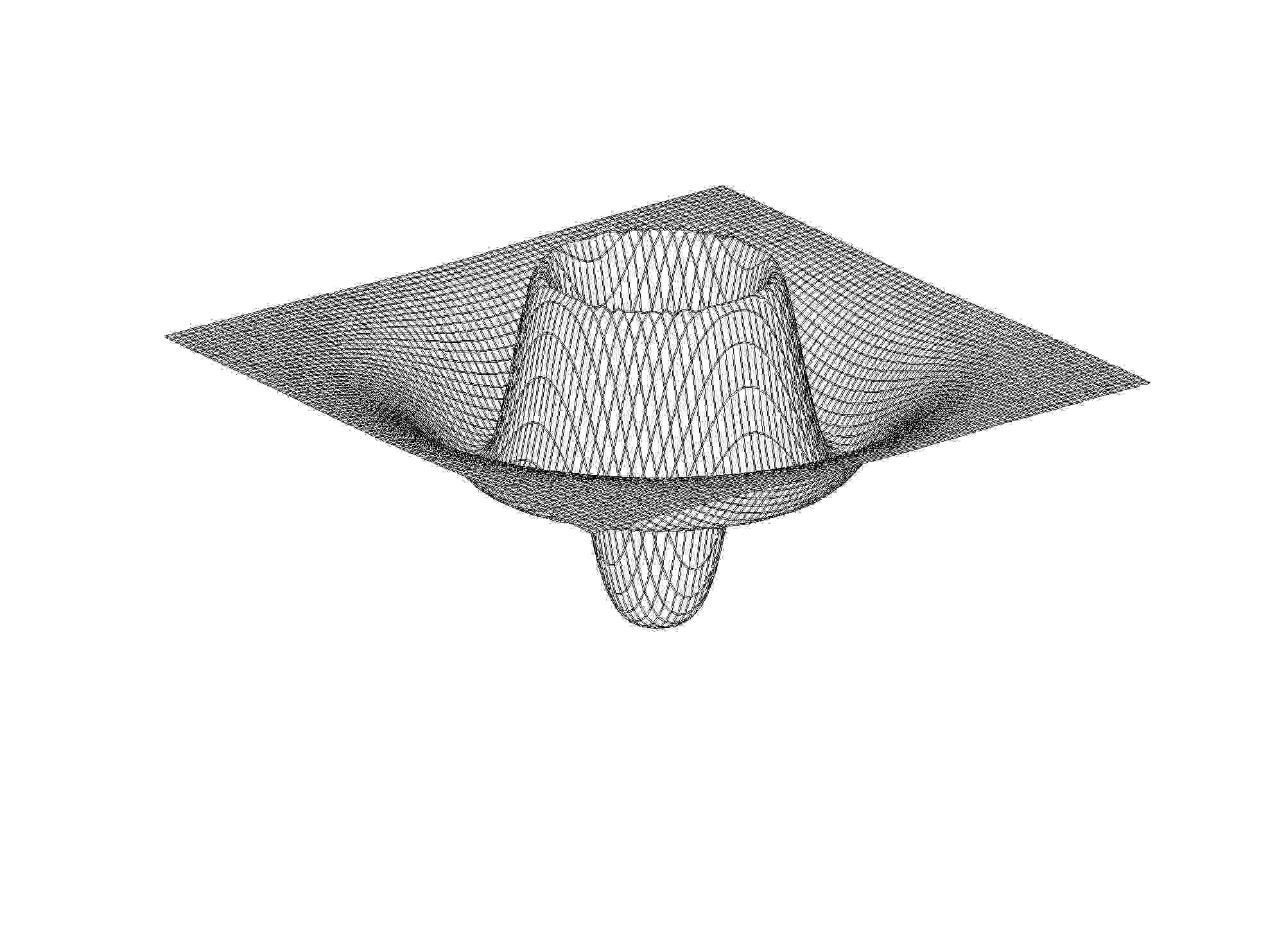}&
d\includegraphics[scale=0.3]{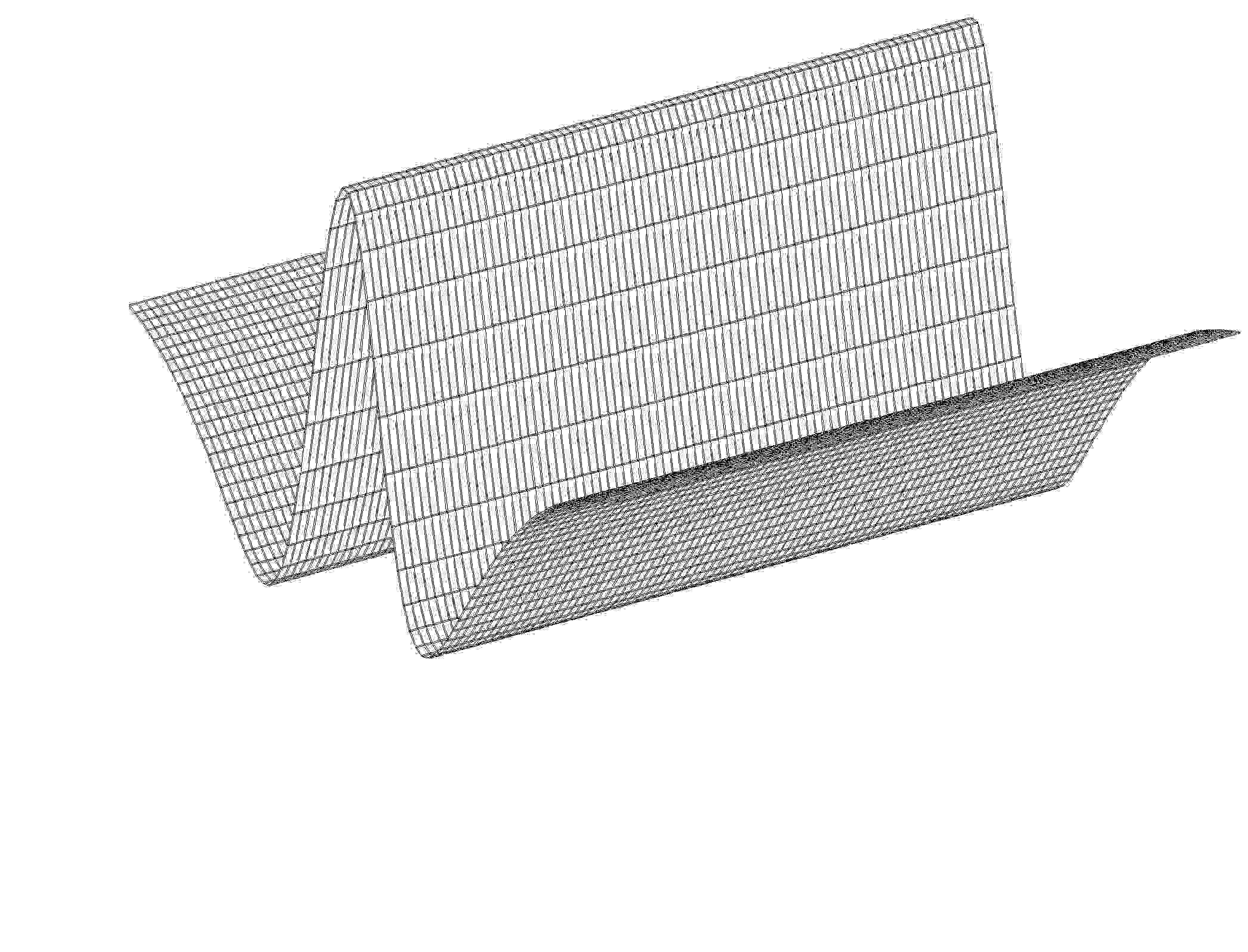}\tabularnewline
\end{tabular}\par\end{centering}

\caption{\label{cap:1} (a) Stable one-bump solution of the neural field equation.
Parameters are $K=2.5$, $k=5$, $M=0.5$, $m=0.5$, $h=-0.281$ (b)
Unstable one-bump solution for the parameter values $K=2.5$, $k=5$,
$M=0.5$, $m=0.5$, $h=-0.294$ (c) Annular solution with the parameters
$K=2.5$, $k=5$, $M=0.5$, $m=0.5$, $h=-0.115$ (d) A part of a
stable stripe-like solution. Parameters are the same as in (c). At
the same parameters also an unstable solution exists (not shown).}
\end{figure}

It has previously been shown that (\ref{eq: Amari eq}) admits rotationally
invariant stationary solutions with disc-shaped activated region (one-bump
solutions). Generically, these arise in the course of a {}``blue
sky bifurcation'' (Strogatz, 1994): no solution is present in the
subcritical parameter region whereas two solution branches bifurcate
as control parameter exceeds the critical value. The two solutions
are rotationally invariant one-bumps. Stability analysis of these
states with respect to rotationally invariant perturbation is essentially
equivalent to stability analysis of one-bump solution of the one-dimensional
model. It reveals that the unstable branch generically corresponds
to the bump of smaller radius. Upon destabilization the region of
activation expands as the solution approaches the stable branch, corresponding
to the circular bump of larger radius (See Fig.~\ref{cap:1}).

Seeking a stationary solution $\overline{u}$ of the two-dimensional
model (\ref{eq: Amari eq}) assuming rotational invariance of the
field (i.e.~$\overline{u}\left(\mathbf{r}\right)\equiv\overline{u}\left(r\right)$
with $r=\left\Vert \mathbf{r}\right\Vert $) leads to a problem, essentially
equivalent to that of finding stationary states in the one-dimensional
model (see Appendix). Two types of solutions besides the circular
one-bumps are readily constructed: solutions with annular activated
regions and solutions with stripe-shaped region of activation (see
Fig.~\ref{cap:1}). The possibility of existence of the former has
been pointed out previously, whilst the latter can be seen as degenerate
annuli of infinite inner radius.

We now turn to the analysis of the stability properties of the above-mentioned
stationary states. Consider the one-bump solution with disc-shaped
activated region (i.e.~$u\left(r\right)>0$ iff $r<R$ ). Consider
dynamics of a small perturbation $\epsilon\eta$, i.e. \begin{equation}
u\left(\mathbf{r},t\right)=\overline{u}\left(\left\Vert \mathbf{r}\right\Vert \right)+\epsilon\eta\left(r\mathbf{,}t\right)\label{eq: perturbation}\end{equation}
 Inserting into (\ref{eq: Amari eq}) and keeping terms of order at
most 1 in $\epsilon$ one finds that to linear order the dynamics
of the perturbation obeys \begin{equation}
\frac{\partial\eta\left(\mathbf{r},t\right)}{\partial t}=-\eta\left(\mathbf{r},t\right)+\int_{\mathbb{R}^{2}}w\left(\left\Vert \mathbf{r}-\mathbf{r'}\right\Vert \right)\delta\left(\overline{u}\left(\mathbf{r'}\right)\right)\eta\left(\mathbf{r'},t\right)d\mathbf{r'}\label{eq:Pert dyn}\end{equation}
 where $\delta\left(\overline{u}\left(\mathbf{r'}\right)\right)$
is Dirac delta function (see Appendix). Substituting an Ansatz of
the form $\eta\left(\mathbf{r},t\right)=\exp\left(\lambda t\right)\xi\left(\mathbf{r}\right)$
one arrives at an eigenvalue problem which in polar coordinates becomes
(see Appendix) \begin{equation}
\lambda\xi\left(r,\theta\right)=-\xi\left(r,\theta\right)+\Gamma\int_{0}^{2\pi}g\left(\theta-\theta'\right)\xi\left(r,\theta'\right)d\theta'\label{eq: Eigenval. problm}\end{equation}
 Here $g$ is a $2\pi$ periodic function depending on kernel $w\left(\left\Vert \mathbf{r}-\mathbf{r'}\right\Vert \right)$
and $\Gamma$ is a constant given by \begin{equation}
\Gamma=R\left|\left.\frac{\partial\overline{u}\left(r\right)}{\partial r}\right|_{r=R}\right|^{-1}\label{eq: Gamma}\end{equation}
 i.e.~the ratio of the radius of activated region $R$ to the absolute
value of the slope of the radial profile of the stationary solution
at $r=R$. Clearly, explicit evaluation of $\Gamma$ requires calculating
the stationary solution which is implicitly given by (\ref{eq: Stat states})
in terms of a double integral. Equation (\ref{eq: Eigenval. problm})
is known as Fredholm's integral equation of the second kind. The integral
operator in the right-hand side of (\ref{eq: Eigenval. problm}) is
compact, bounded and self-adjoint implying that every spectral value
is an eigenvalue, all eigenvalues are real, each eigenspace is finite-dimensional
and zero is the only possible accumulation point of eigenvalues (Kreyszig,
1978). Solving (\ref{eq: Eigenval. problm}) we obtain eigenvalues
and eigenfunctions which in polar coordinates read (see Appendix)
\begin{eqnarray}
 & \lambda_{n}=-1+\Gamma\int_{0}^{2\pi}g\left(\theta\right)\cos\left(n\theta\right)d\theta\nonumber \\
 & \xi_{n}=\int_{0}^{2\pi}w\left(r,\theta,R,\theta'\right)\cos\left(n\theta'\right)d\theta'\label{eq: Eigenvals}\end{eqnarray}
 where $R$ is the radius of the circular activated region. The $n$th
eigenfunction is $2\pi n$-periodic in $\theta$, implying that it
is $D_{n}$-symmetric (symmetry with respect to rotation by $2\pi/n$
around the origin and with respect to reflections on the $n$ respective
symmetry planes; the shape in Fig.~\ref{cap:colorf2}a, e.g., is
$D_{4}$-symmetric) and corresponds to a multi-periodic deformation
of a circle. Eigenvalue spectra for one-bump solutions are given in
Fig.~\ref{cap:2}. Certain features of these are readily interpretable.
E.g.~we see that for the bump of smaller radius the eigenvalue $\lambda_{0}$,
corresponding to rotationally invariant deformation, is positive implying
instability with respect to perturbations in radius of the bump. The
spectrum, corresponding to the larger bump, however, is non-\textcolor{black}{positive},
implying stability with respect to arbitrary perturbation in agreement
with earlier results. Also, the eigenvalue $\lambda_{1}$, that corresponds
to a $2\pi$-periodic deformation (or, equivalently, to a translation
of the bump) vanishes, reflecting the translational invariance of
(\ref{eq: Amari eq}). Exploiting this observation, it appears possible
to re-express $\Gamma$ in (\ref{eq: Gamma}) more explicitly as $\Gamma=1/\int_{0}^{2\pi}g\left(\theta\right)\cos\left(\theta\right)d\theta$,
whereby the expressions for the other eigenvalues simplifies to \begin{equation}
\lambda_{n}=-1+\frac{\int_{0}^{2\pi}g\left(\theta\right)\cos\left(n\theta\right)d\theta}{\int_{0}^{2\pi}g\left(\theta\right)\cos\left(\theta\right)d\theta}\label{eq: Eigenvals of R}\end{equation}

\begin{center}%
\begin{figure}[H]
\begin{centering}\begin{tabular}{ccc}
a\includegraphics[scale=0.21]{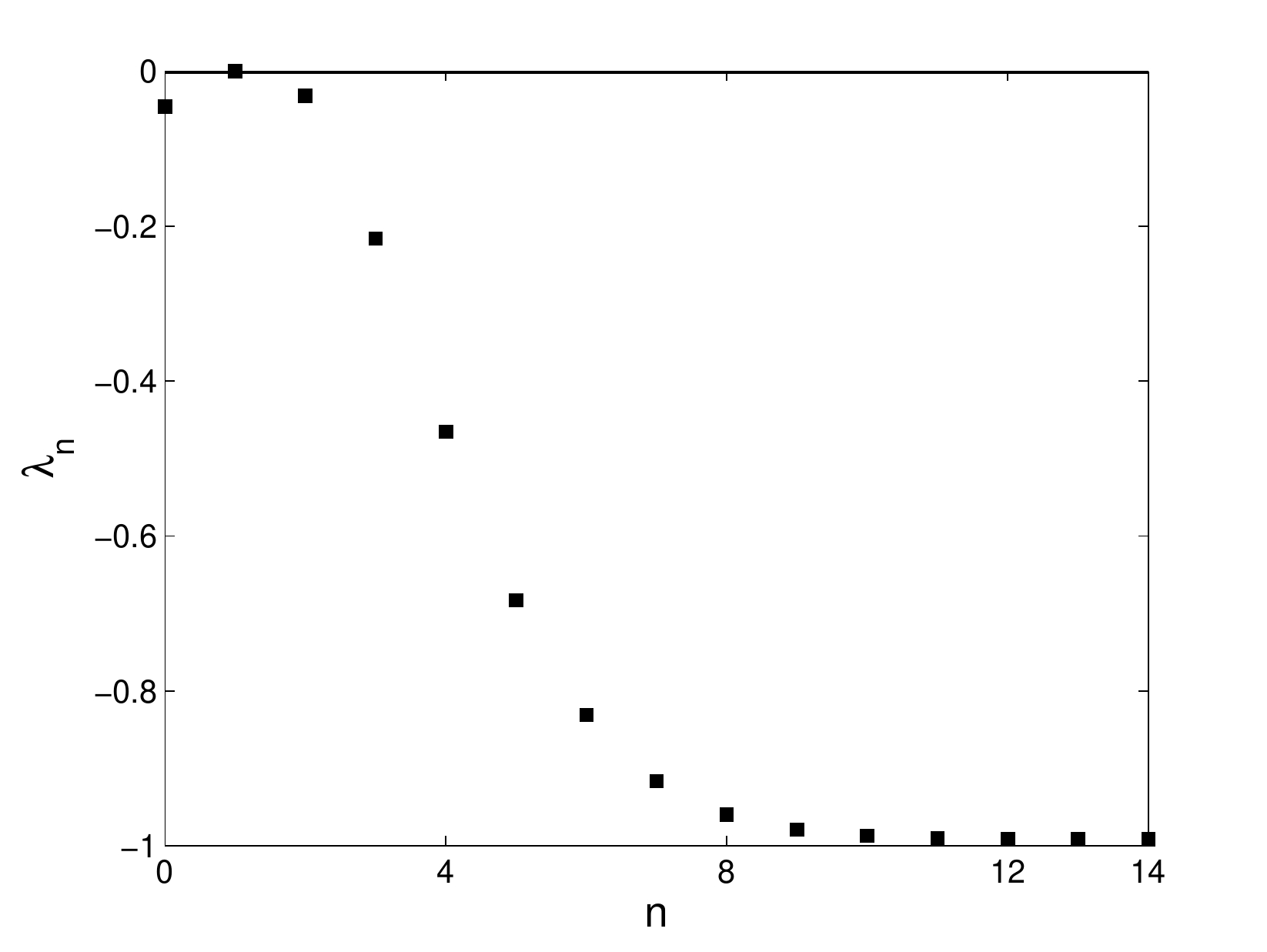}&
b\includegraphics[scale=0.21]{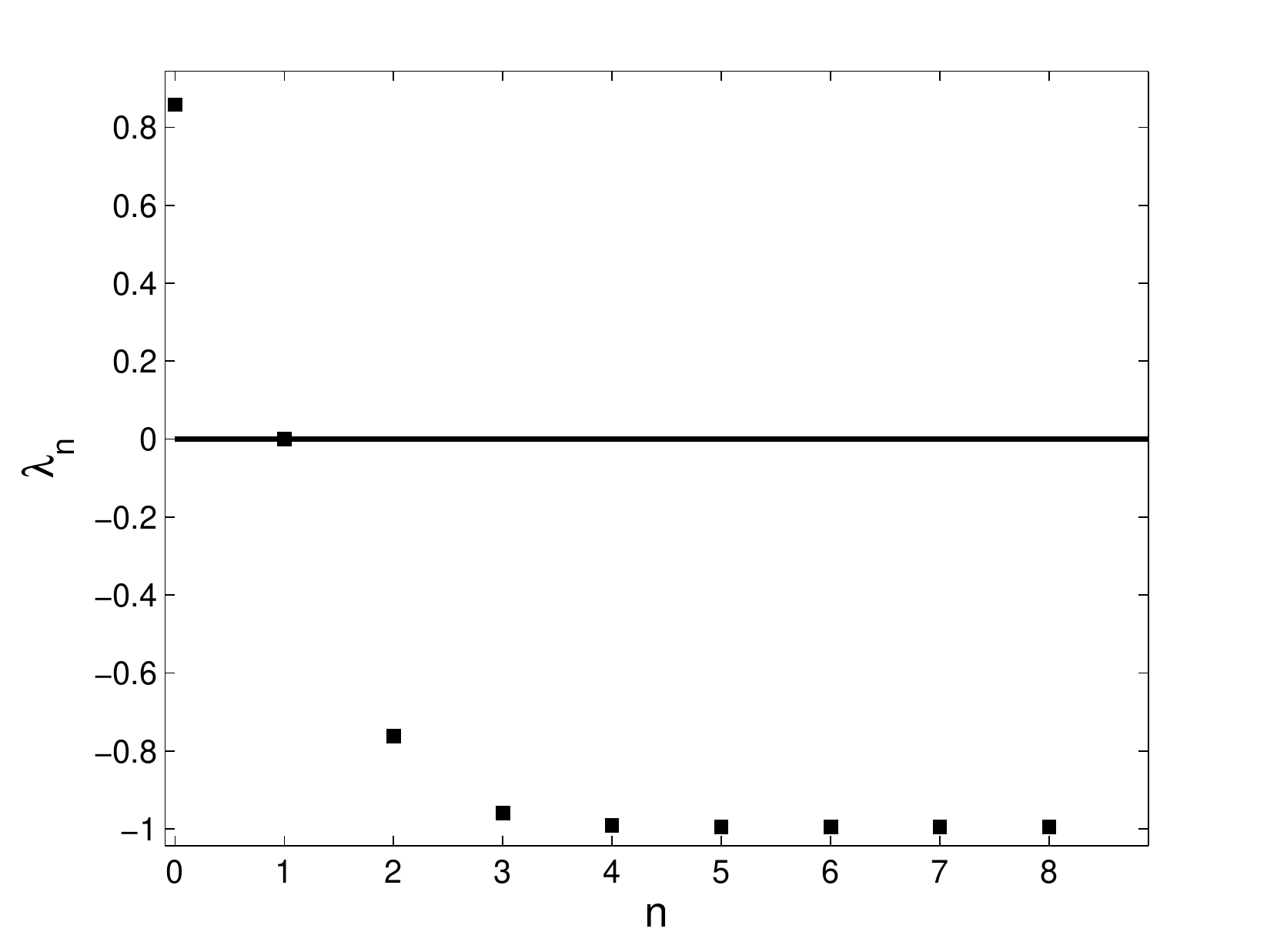}&
c\includegraphics[scale=0.21]{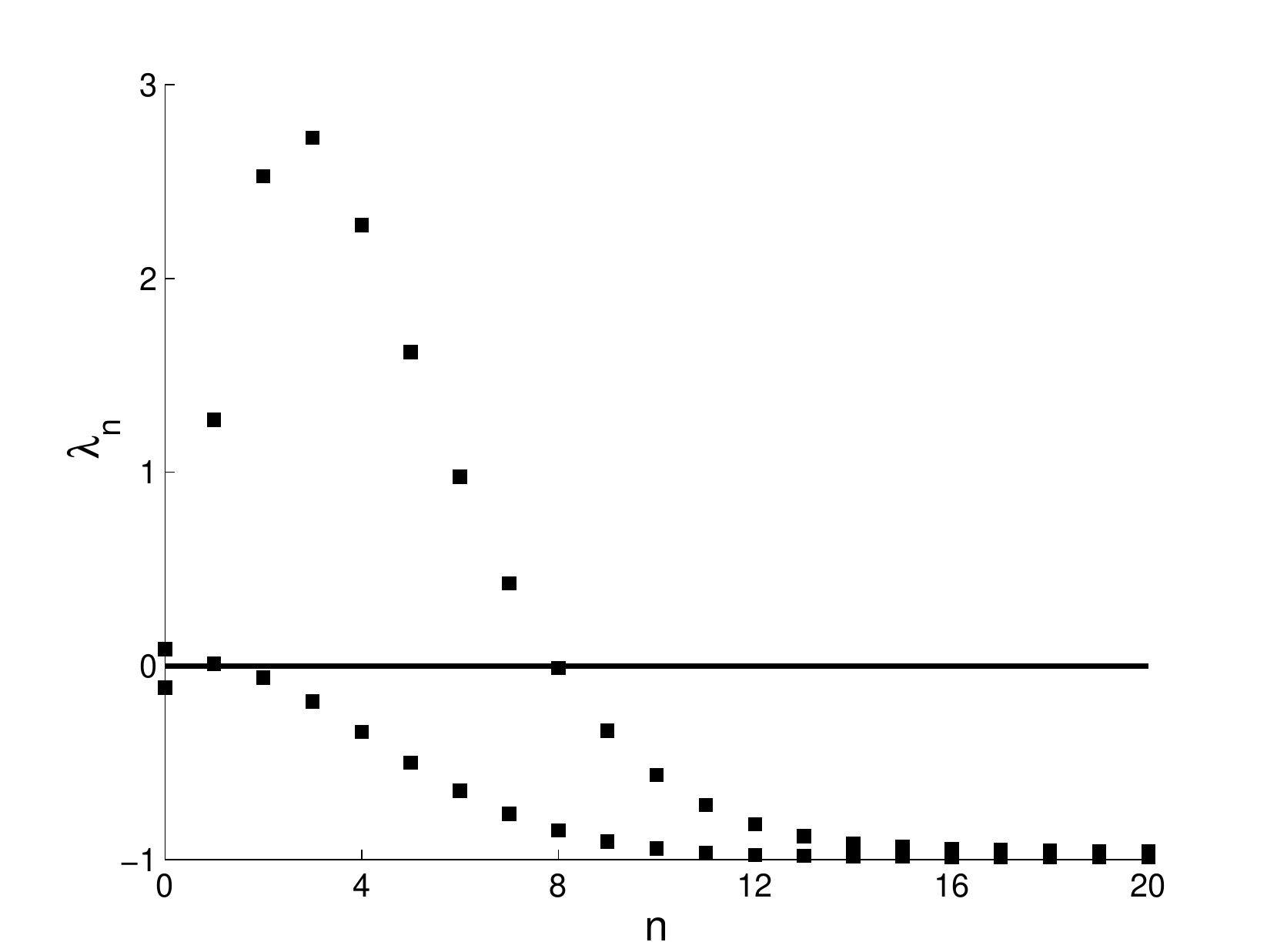}\tabularnewline
\end{tabular}\par\end{centering}

\caption{\label{cap:2} Spectra of \textcolor{black}{corresponding} to solutions
in Fig.~\ref{cap:1} a, b, and c, respectively. The positivity of
the zeroth mode in (b) indicates the instability of this solution
with respect to circular perturbations. The solution (a) is stable,
with a marginal instability with respect lateral shift as visible
from the vanishing first eigenvalue. The instability of the annulus
solution in (c) shows a characteristic scale which causes the solution
to split into three single bumps.}
\end{figure}
\par\end{center}

Note that contrary to (\ref{eq: Eigenvals}), Eq.~\ref{eq: Eigenvals of R}
does not contain explicitly the stationary solution $\overline{u}$,
which allows us to calculate the $n$th eigenvalue as a function of
the radius of activated region without evaluating double integrals
in the implicit expression for $\overline{u}$. Only integrals over
one-dimensional manifolds appear in (\ref{eq: Eigenvals of R}), greatly
simplifying the calculation of the spectrum. Apart from theoretical
considerations, this is of importance for technical applications since
the knowledge of stability properties of solutions of (\ref{eq: Amari eq})
could affect their use for representing probability distributions
e.g.~in implementations of autonomous robot memory.

\begin{figure}[H]
 \centerline{ a\includegraphics[scale=0.33]{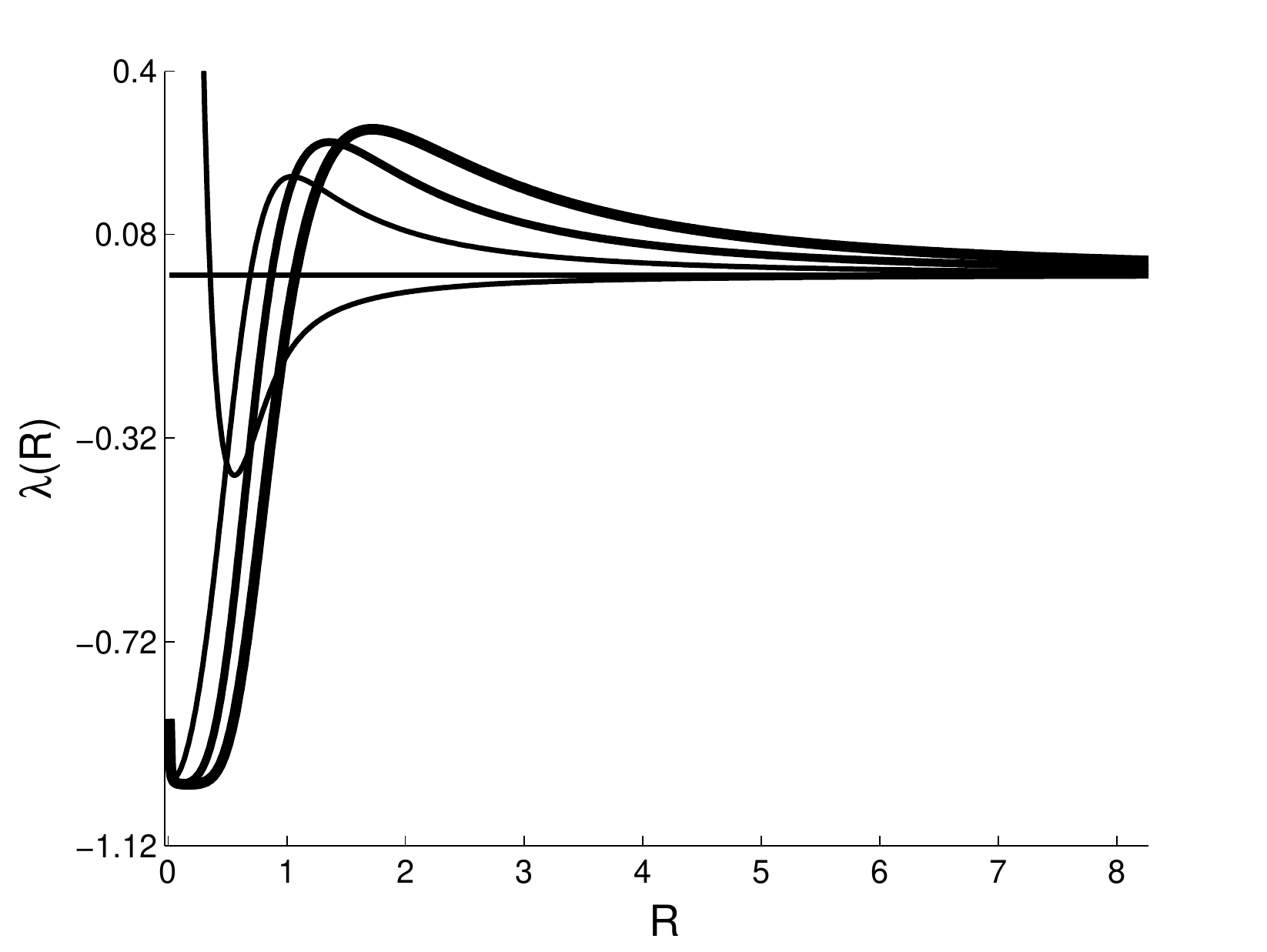} \hskip 0.3cm
b\includegraphics[scale=0.33]{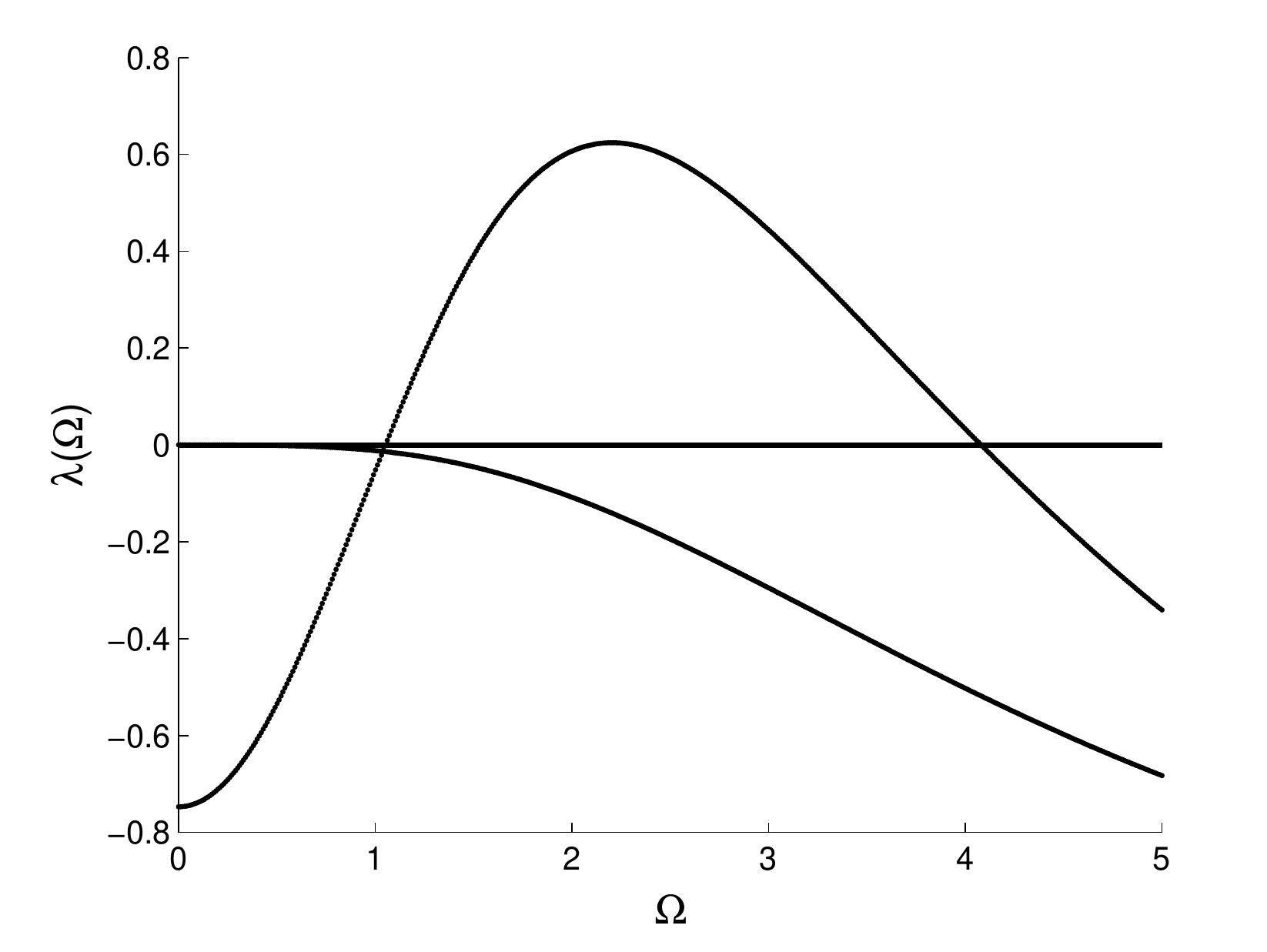}}

\caption{\label{cap:3} (a) Four eigenvalues versus bump radius $R$. The
curve which represents $\lambda_{0}$ described the stability with
respect to perturbations in bump radius. The others are (from left
to right) $\lambda_{2}$ (reflection-invariant deformation), $\lambda_{3}$
($D_{3}$-invariant eigenmode) and $\lambda_{4}$ ($D_{4}$-invariant
eigenmode). $\lambda_{1}$ is identically zero reflecting the metastability
of the solutions w.r.t.~\textcolor{black}{lateral} shifts. Parameters
are: $K=1.5$, $k=5$, $M=0.5$, $m=1.5$ (b) Spectrum determining
stability of stripe-shaped solutions shown in Fig.~\ref{cap:1}(d).
Only the information for the stable solution if given here. This solution
looses stability for a band of values of $\Omega$. }
\end{figure}

Previous work (Werner-Richter 2001) conjectured the bifurcation branch
corresponding to the stable one-bump to remain stable for all values
of control parameter. Using (\ref{eq: Eigenvals of R}) this is readily
proved false: higher and higher frequency eigenmodes progressively
turn unstable as the radius of the stationary solution is increased
(see Fig.~\ref{cap:3}).

\begin{figure}[H]
\begin{centering}\begin{tabular}{cccc}
\includegraphics[scale=0.16]{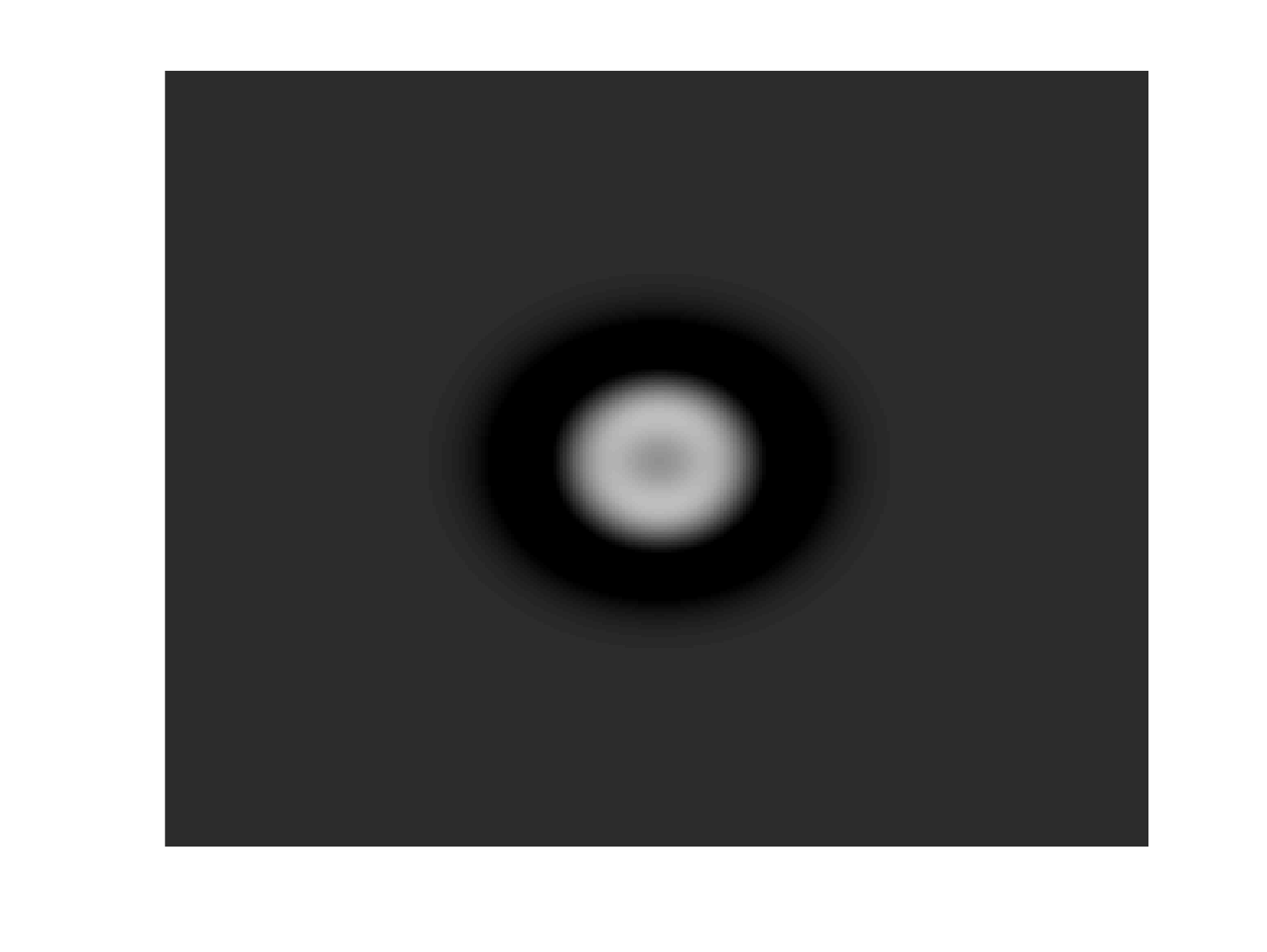}&
\includegraphics[scale=0.16]{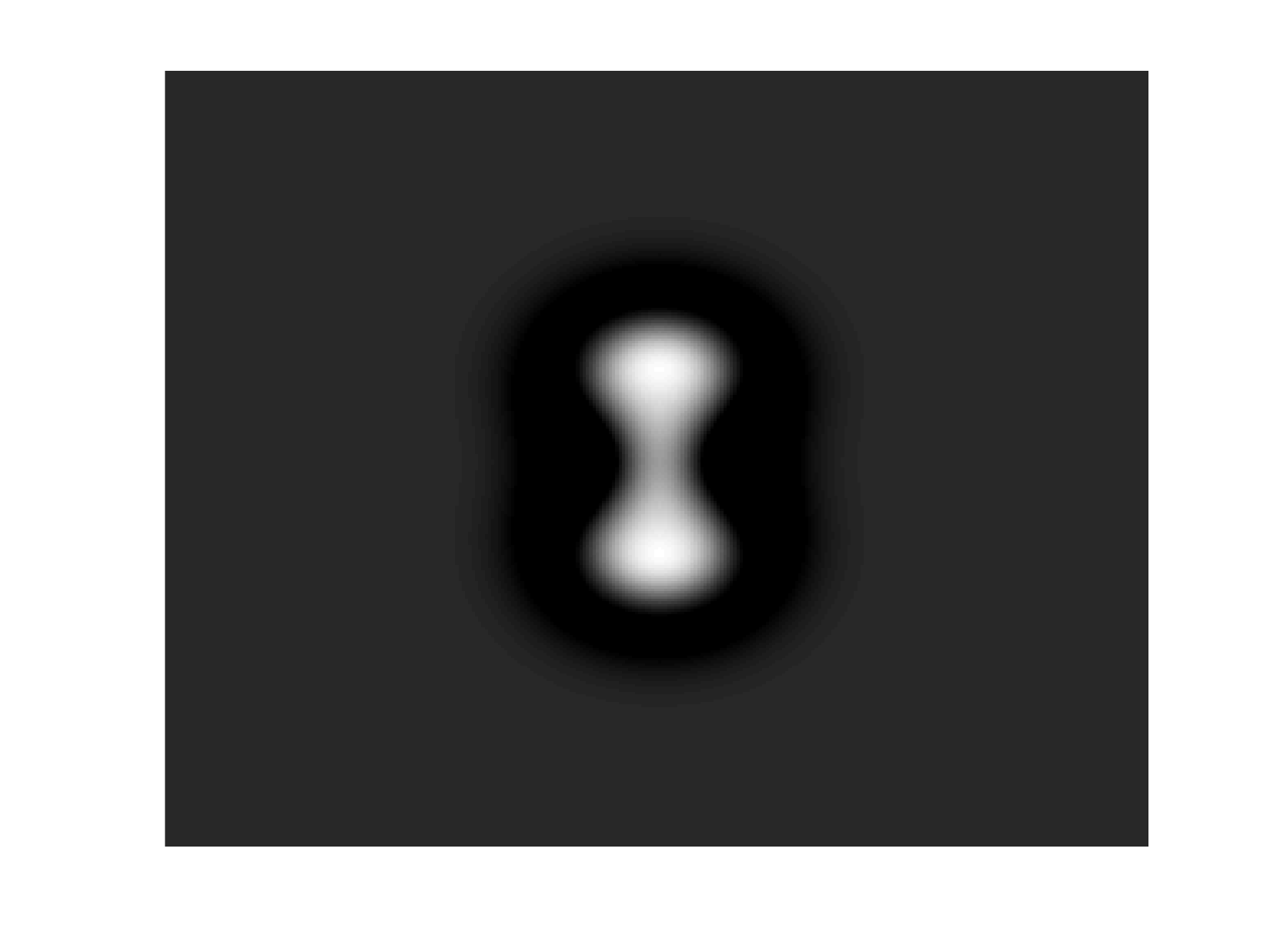}&
\includegraphics[scale=0.16]{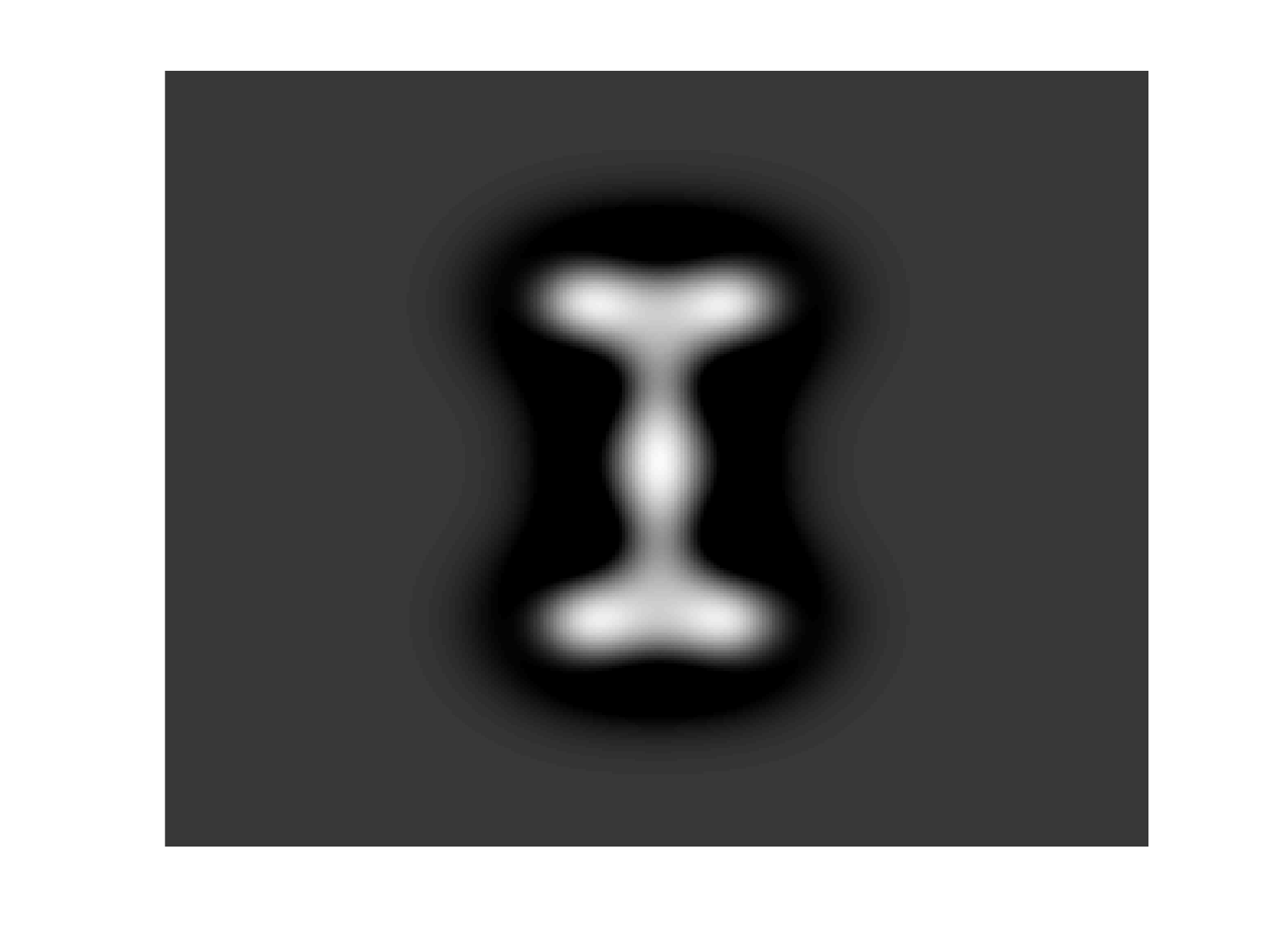}&
\includegraphics[scale=0.16]{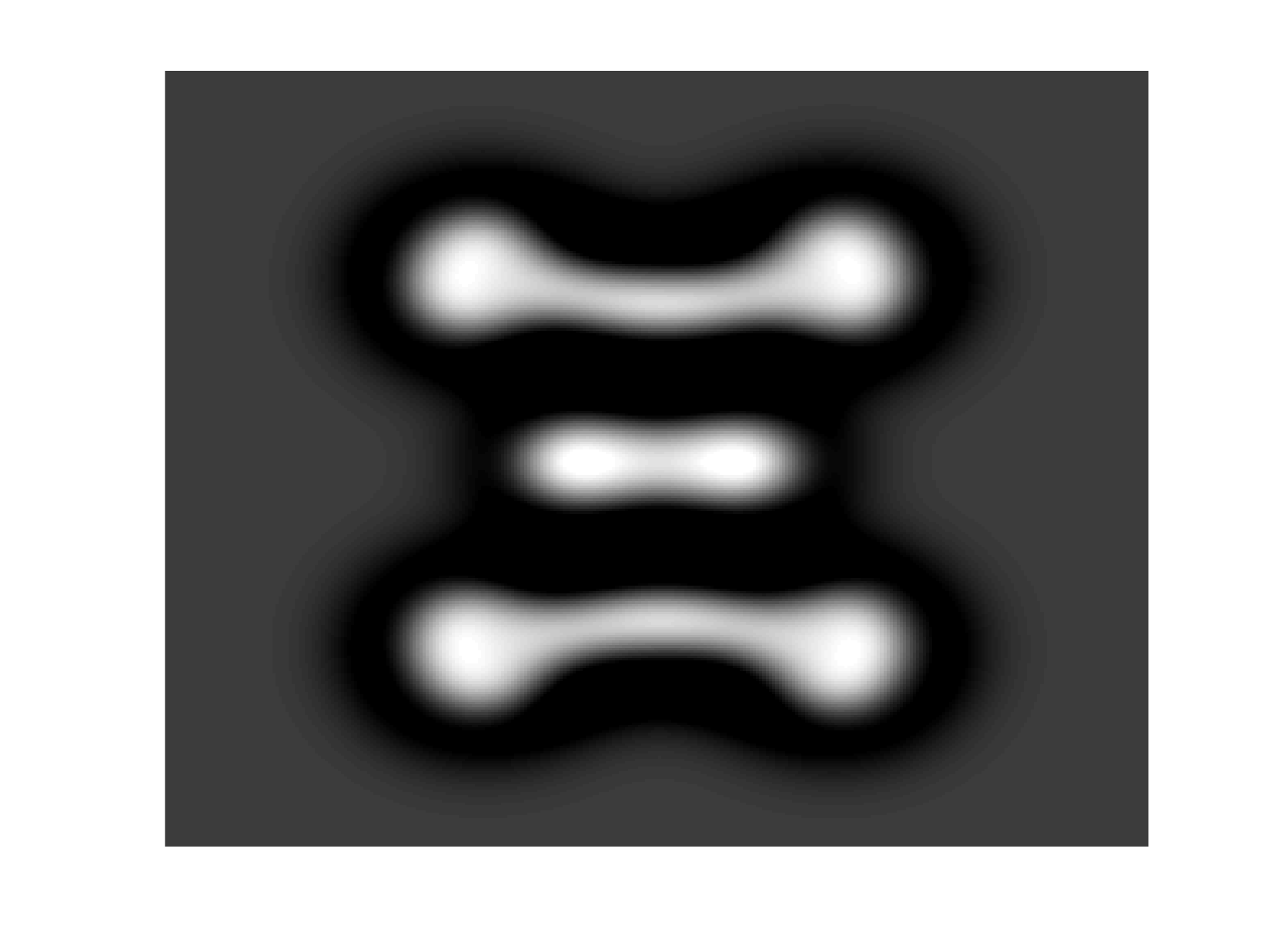}\tabularnewline
\includegraphics[scale=0.16]{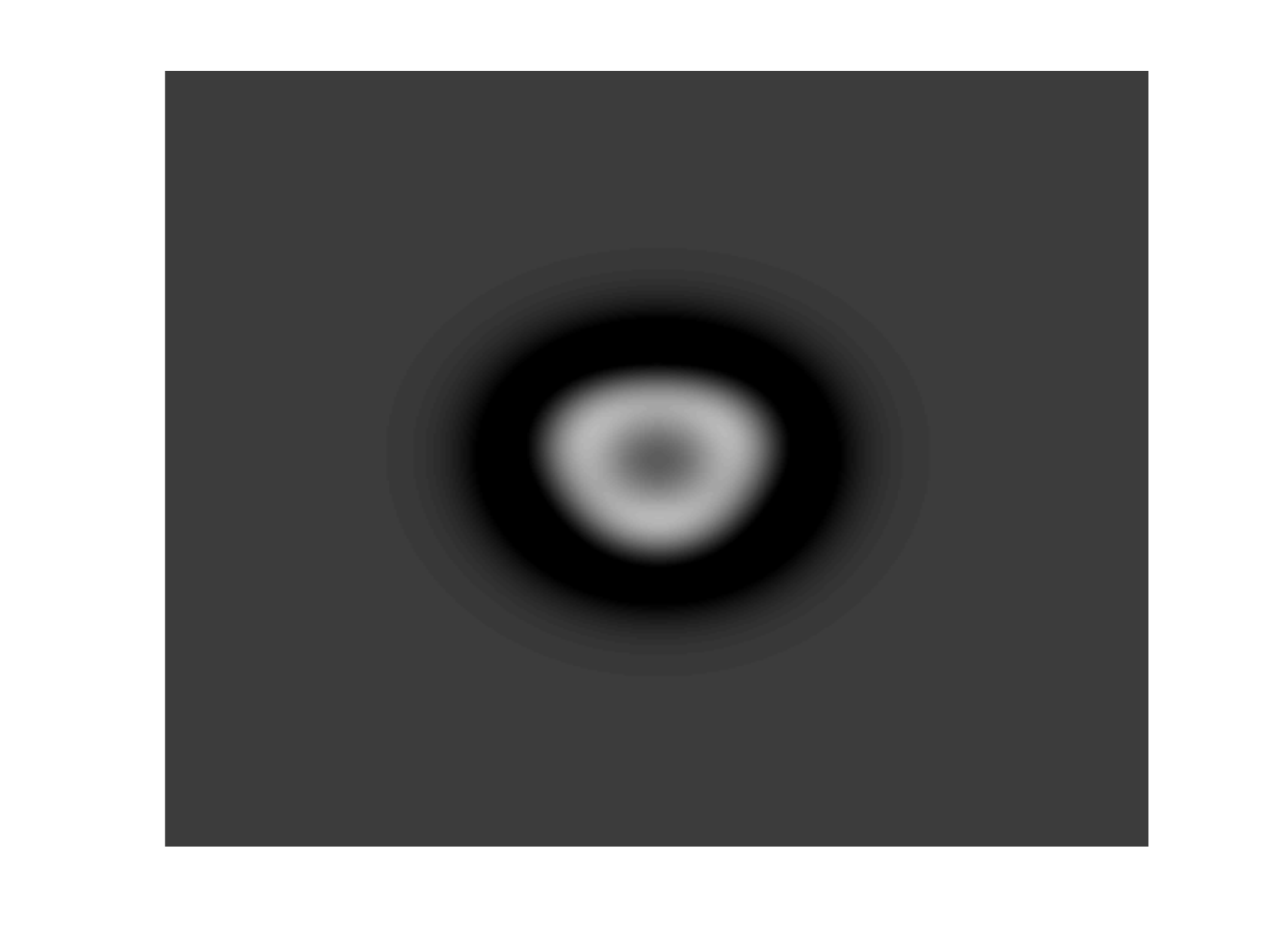}&
\includegraphics[scale=0.16]{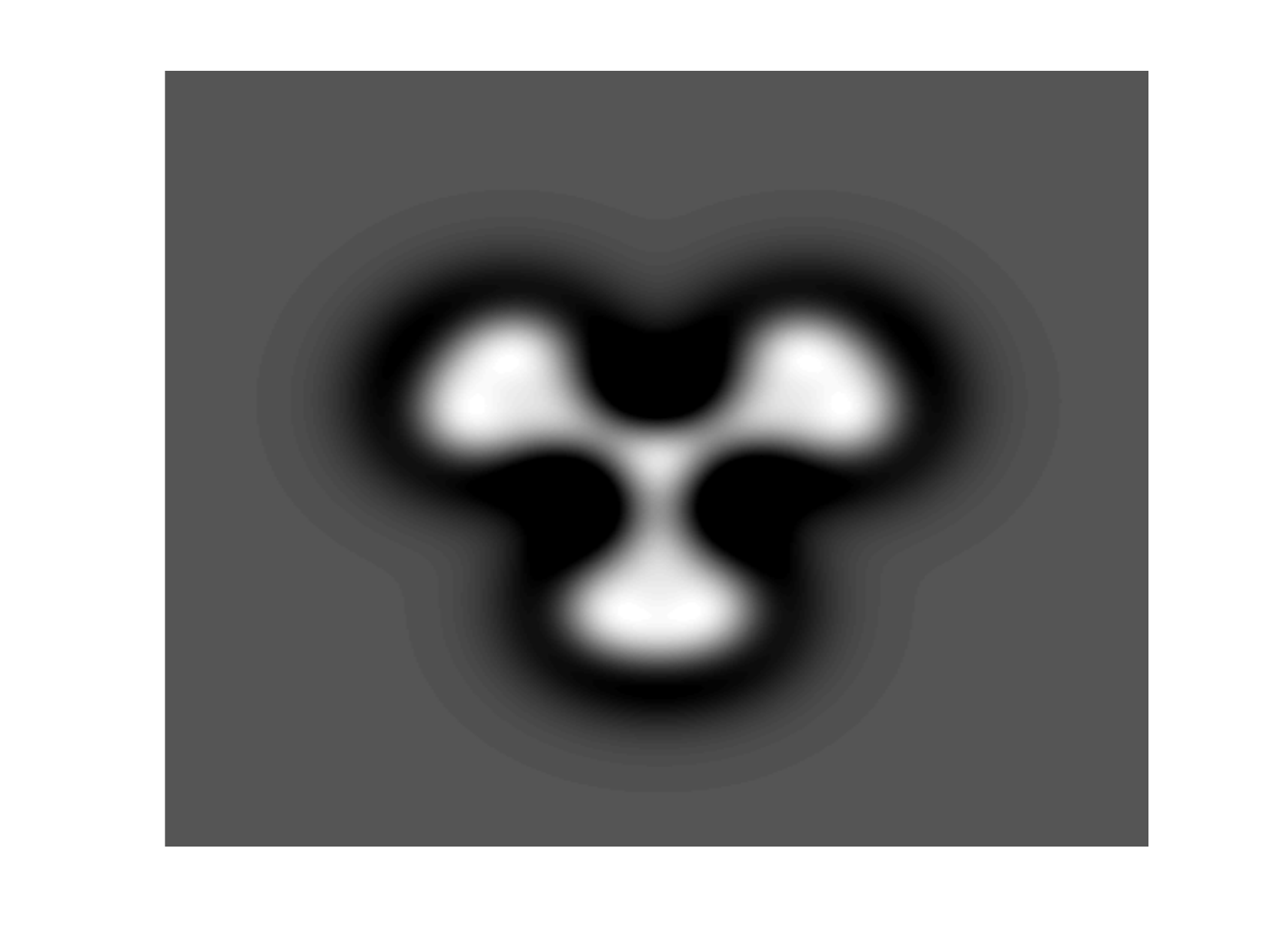}&
\includegraphics[scale=0.16]{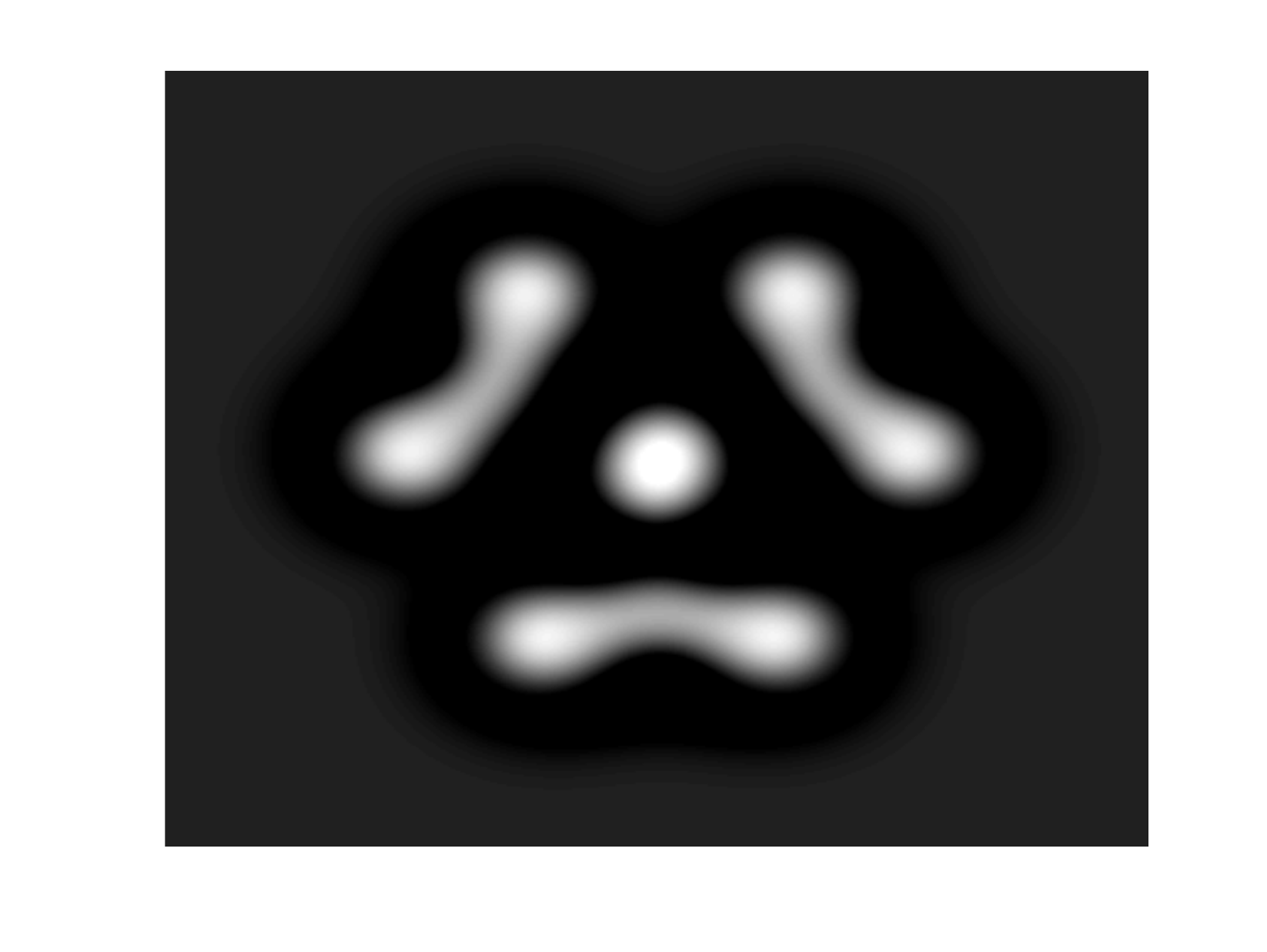}&
\includegraphics[scale=0.16]{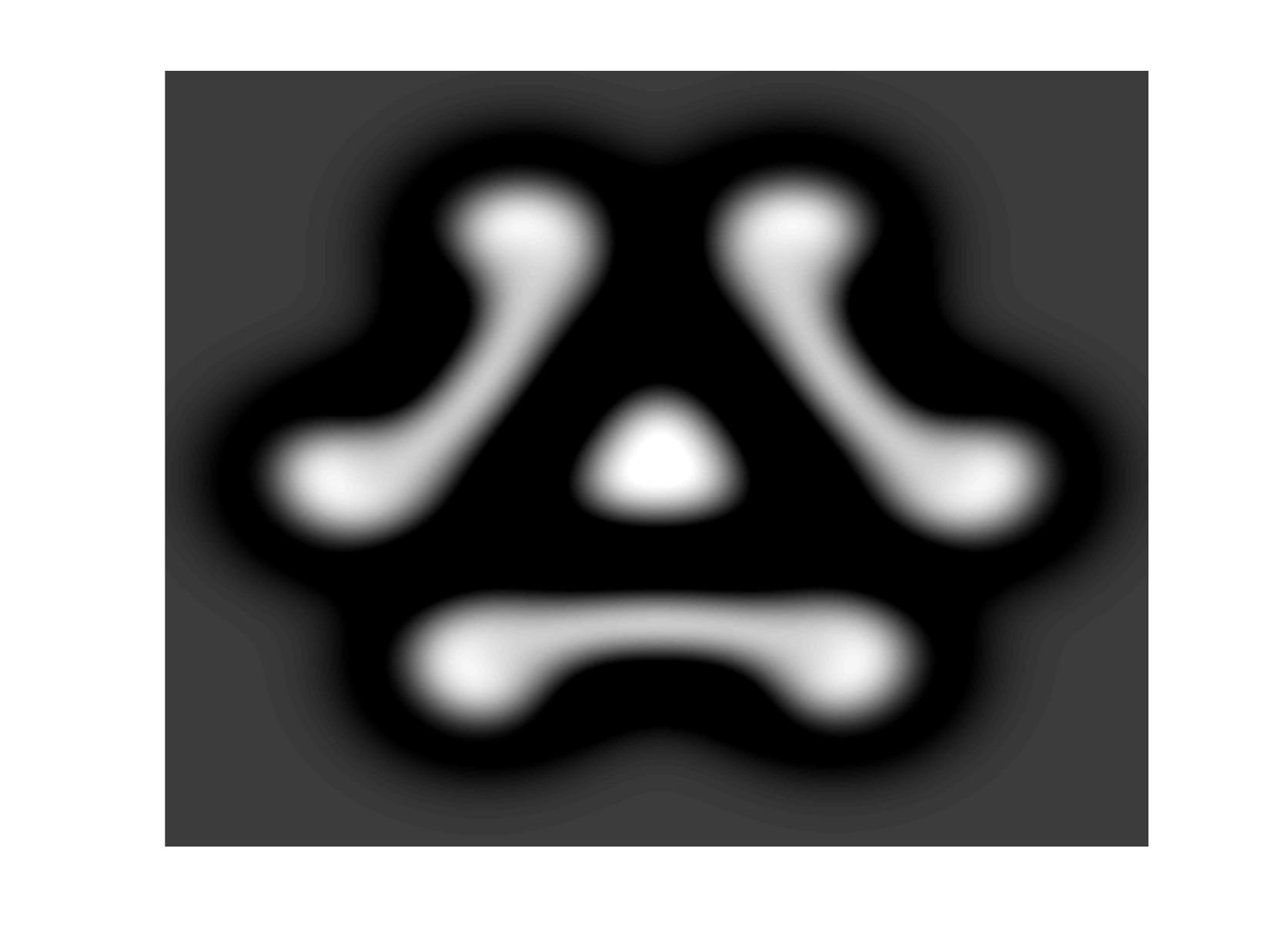}\tabularnewline
\end{tabular}\par\end{centering}

\caption{\label{cap:4} (\textcolor{black}{U}pper panel) Time evolution
of an unstable bump, undergoing stability loss at a $D_{2}$-eigenmode.
Parameters are: $K=1.5$, $k=5$, $M=0.5$, $m=1.5$, $h=-1.46\cdot10^{-2}$.
(\textcolor{black}{L}ower panel) Time evolution of an unstable bump,
undergoing stability loss at a $D_{3}$ invariant eigenmode. Parameters:
$K=1.5$, $k=5$, $M=0.5$, $m=1.5$, $h=-4.43\cdot10^{-3}$.}
\end{figure}

\begin{figure}[H]

\begin{centering}\begin{tabular}{cc}
a\includegraphics[scale=0.3]{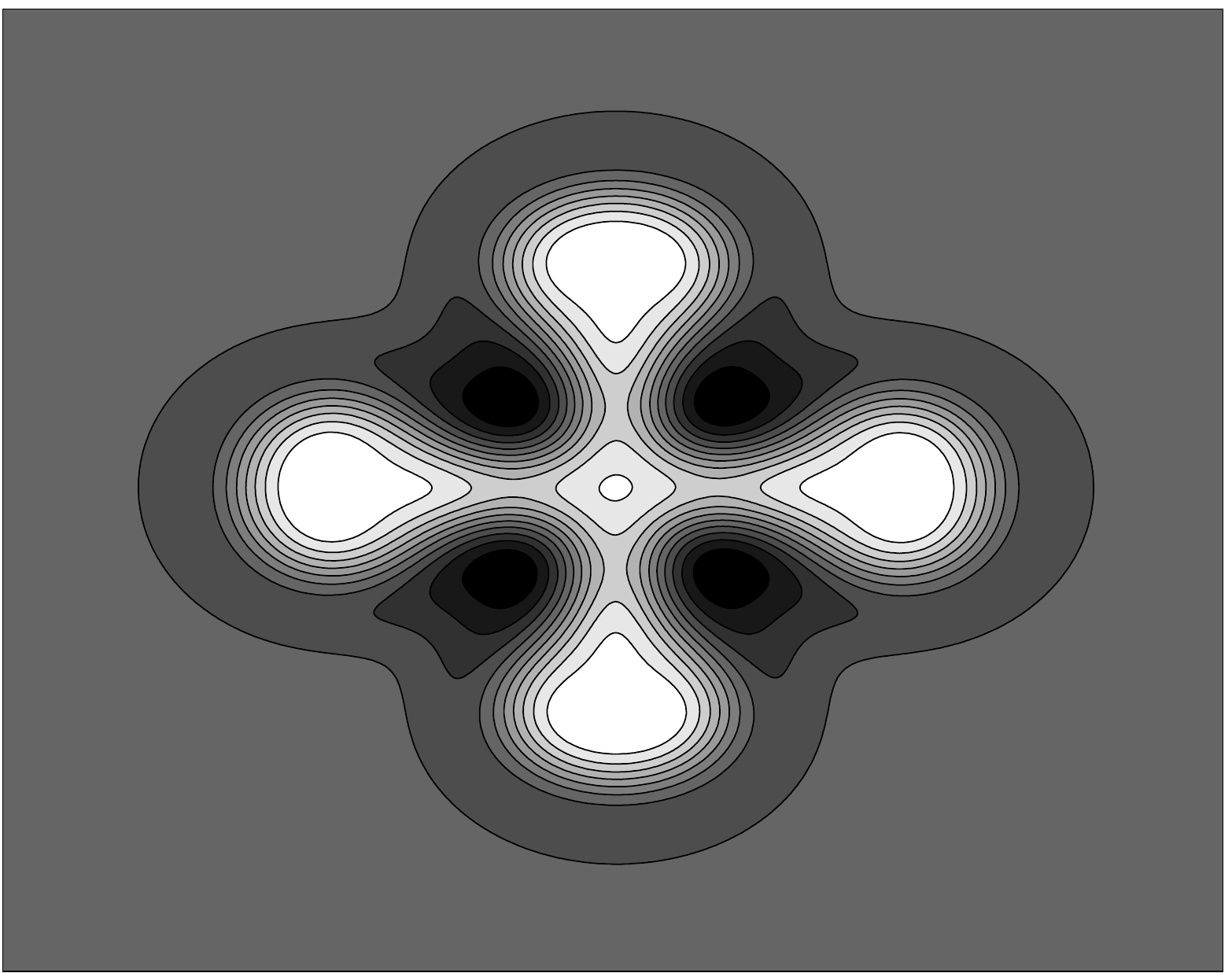}&
b\includegraphics[scale=0.3]{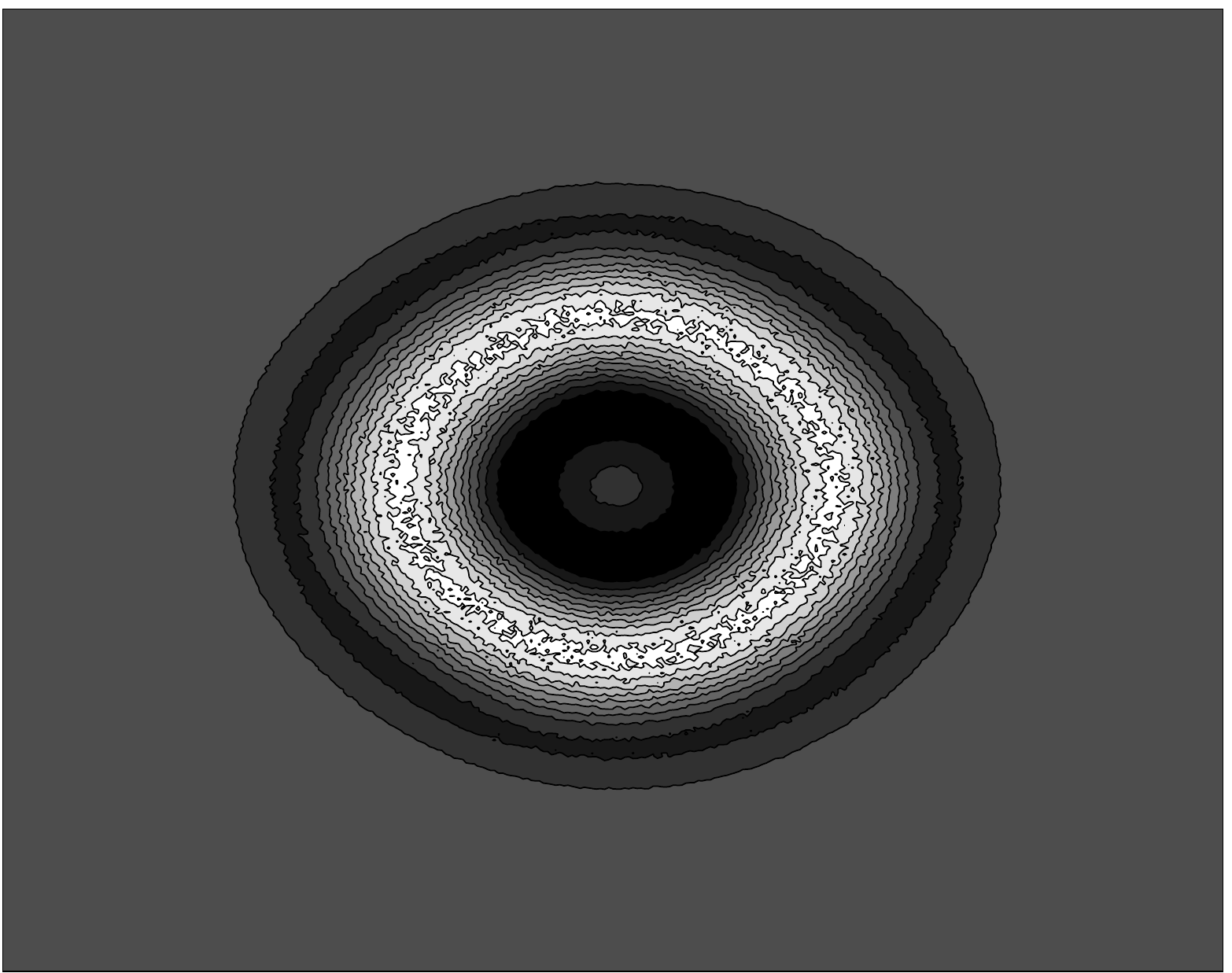}\tabularnewline
c\includegraphics[scale=0.3]{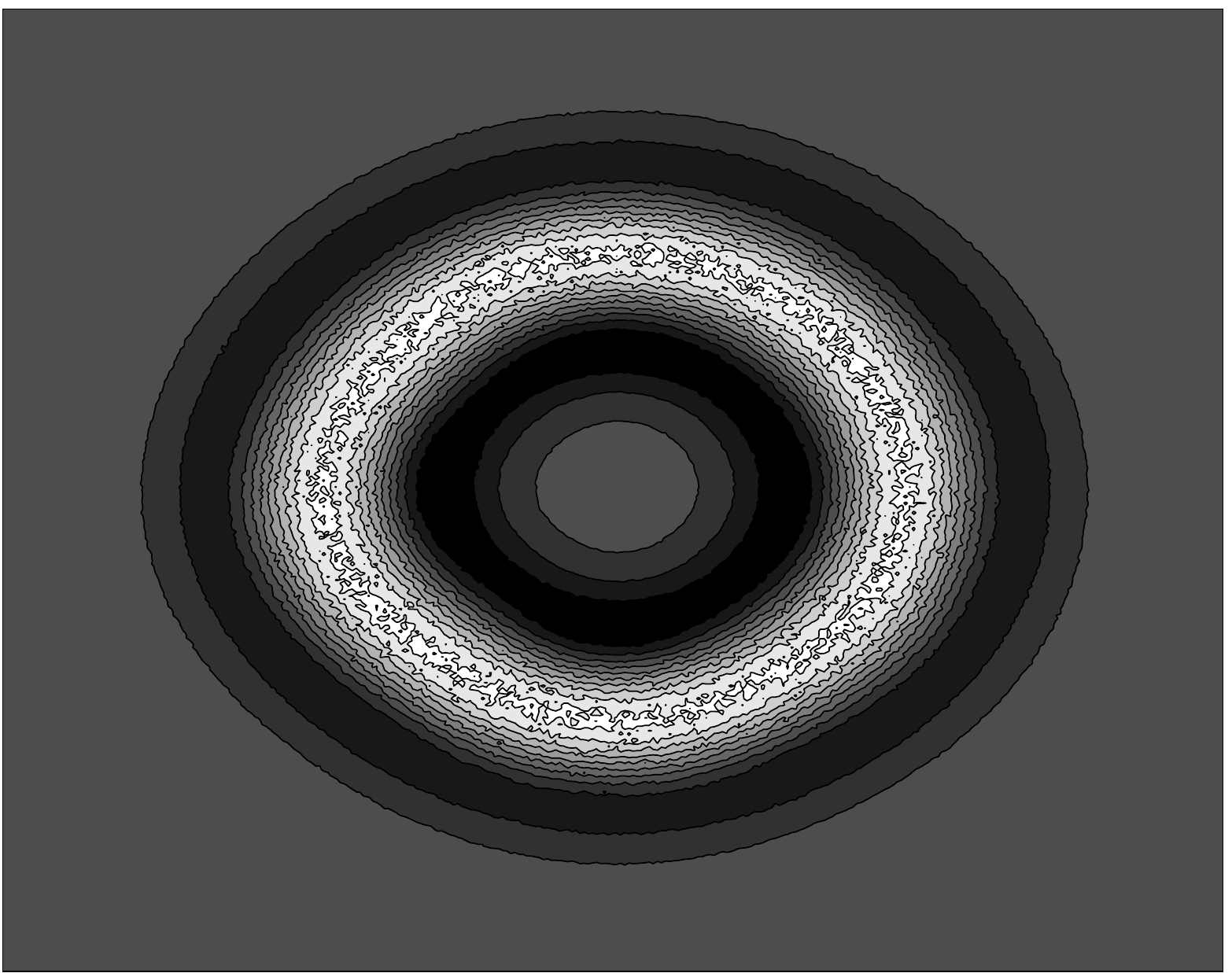}&
d\includegraphics[scale=0.3]{noise_3.pdf}\tabularnewline
\end{tabular}\par\end{centering}

\caption{\label{cap:colorf2} Dynamics of the modified field with and without
noise. (a) In the absence of noise an unstable bump develops four
identical protrusions subsequently splitting into four separate blobs.
(b) In presence of noise a ring-shaped region of activation is initially
annular, but (c and d) develops gradually into an irregularly meandering
profile. Parameters are $k=5$, $m=1.6$,$K=1.5$, $M=0.45$, $h=-0.0505$,
$p=-6.25\cdot10^{-4}$.}
\end{figure}

Strictly speaking, the assertion that linear stability analysis as
outlined above correctly determines stability properties of stationary
solutions relies on additional assumptions on operators appearing
in the right-hand side of (\ref{eq: Amari eq}) (Schaefer and Golubitsky
1988). For infinitely-dimensional non-smooth fields these generically
need not hold. In order to check whether stability analysis is indeed
adequate, correctly determining stability properties of the stationary
solutions, we performed a number of numerical simulations. Fig.~(\ref{cap:4})
depicts a simulation of unstable one-bump whose corresponding stability
spectrum reveals that the maximal (positive) eigenvalue is that of
the $D_{2}$-symmetric eigenmode. As the time elapses, initially circular
activated region keeps deforming, forming blob-like protrusions. These
subsequently bud off from the middle-bump. The newly-formed activated
domains keep splitting, progressively tiling the plane in a hexagonal
pattern. We would like to stress that the pattern which is formed
immediately after destabilization of the stationary state is $D_{2}$-symmetric
as would be expected from the properties of the eigenvalue spectrum.
That is, stability properties of the stationary solutions as well
as qualitative aspects of the pattern\textcolor{black}{,} forming
upon destabilization of the steady state\textcolor{black}{,} are
correctly predicted from the above-mentioned linear stability analysis.

Symmetry breaking accompanying destabilization of a stationary bump
was examined for a broad range of parameters (e.g.~parameters which
yielded solutions with maximal eigenvalue of the respective spectrum
corresponding to $D_{3}$-, $D_{4}$- and $D_{8}$-symmetric perturbations,
see Fig.~\ref{cap:4}). In all of the cases, the course of the symmetry
breaking was appropriately determined by linear stability.

Stability analysis of annular solutions proceeds along the same lines
as that of one-bumps (see Appendix). The essential difference is that
instead of a singe equation (\ref{eq: Eigenval. problm}), a system
of two equations results, meaning that to every non-negative whole
number corresponds a pair of real eigenvalues. (In general, to every
boundary of an activated domain there corresponds an equation in the
corresponding eigenvalue problem). We find that in the case of annular
solution shown in Fig.~\ref{cap:1}d the largest eigenvalue corresponds
to a $D_{3}$-symmetric perturbation. Simulations demonstrate that
initially rotationally invariant annulus splits into three adjacent
blobs which gradually drift apart (not shown). The symmetry of the
resulting state is the same as that of the largest eigenvalue. Again,
stability of the stationary solution as well as qualitative aspects
of the emerging pattern are readily predicted from the analytically
computed spectrum.

Finally, let us consider the stripe-shaped solutions. Their stability
is governed by an eigenvalue problem, similar to that governing stability
of the annuli. However, the corresponding linear operator is no longer
compact (this is a consequence of activated region being unbounded)
and the spectrum needs no longer remain discrete. Actually, in this
case the spectrum is continuous: to every real corresponds a pair
of (real) eigenvalues. Fig.~\ref{cap:3}b shows results of linear
stability analysis of the stripe solution, depicted in Figure \ref{cap:1}d.
In the corresponding simulation the stripe is seen to split into a
row of separate bumps -- a {}``chain-of-pearls'' configuration.
The inter-bump separation is the same as the wave-length of the eigenmode,
corresponding to the largest eigenvalue. Again, stability and semi-quantitative
properties of patterns resulting from the stationary state destabilization
are readily predicted from the respective eigenvalue spectrum.

In summary, preceding section describes all of the stationary non-homogeneous
solutions of the two-dimensional Amari equation known up to date and
exhaustively examines their stability properties. Quite strikingly,
stability analysis of the non-homogeneous steady state is possible
since the eigenvalue problem (\ref{eq: Eigenvals}) is effectively
one-dimensional although a two-dimensional system is being considered.

\section{Modified equation}

A long-standing question regarding Amari-model is existence of non-rotationally
invariant stationary solutions with bounded and connected region of
activation. Such states are likely to bifurcate from circular solutions
upon destabilization at a non-rotationally invariant eigenvalue. In
fact, stability loss of one branch is always accompanied by emergence
of another in its vicinity provided that the mappings defining the
dynamical system under consideration are sufficiently smooth (Crandall
and Rabinowitz 1971). However, in the above simulations exclusively
periodic patterns resulted upon destabilization of circular solutions.

In order to address existence of non-rotationally invariant solutions
of (\ref{eq: Amari eq}) with bounded and connected activated region
we shall modify the original Amari-model (\ref{eq: Amari eq}) so
as to obtain a related ({}``modified'') equation fulfilling the
following three conditions.

\begin{enumerate}
\item Stationary solutions of the modified equations should be solutions
of the unmodified equation (\ref{eq: Stat states}). 
\item The modified equation should not admit solutions with unbounded activated
region. 
\item Stability of a solution of Eq.~(\ref{eq: Stat states}) should remain
unaltered by the modification. 
\end{enumerate}
According to condition 3 
rotationally invariant stationary solutions behave like those of the
unmodified Amari-model, admitting symmetry breaking at a non-rotationally
invariant eigenvalue. However, the symmetry breaking cannot result
in a spatially extended periodic pattern according to condition 2.
Consequently, a non-rotationally invariant localized state is likely
to emerge. It will be a solution of the original (unmodified) Amari
model with desired properties, provided that its region of activation
remains connected in the course of destabilization. Note, however,
that conditions 1-3 do not suffice to ensure that the activated domain
will not start splitting into separate disconnected region upon symmetry
breaking.

We now turn to the construction of a modification of (\ref{eq: Amari eq})
satisfying the above-mentioned conditions 1-3. Consider some circular
one-bump solution $\overline{u}_{h}$ of (\ref{eq: Amari eq}) with
resting potential $h$ and the area of activated region $A[\overline{u}_{h}]$.
Let us modify (\ref{eq: Amari eq}) according to

\begin{equation}
\partial_{t}u=-u\left(\mathbf{r},t\right)+\int_{R\left[u\right]}\left[w\left(\left|\mathbf{r}-\mathbf{r'}\right|\right)+q\right]d\mathbf{r'}-qA[\overline{u}_{h}]+h'\label{eq: Modified eq, q}\end{equation}
 where $q$ is any real number. Suppose that $\overline{u}$ is a
stationary solution of (\ref{eq: Modified eq, q}) for a certain $q$.
Substituting $\overline{u}$ into (\ref{eq: Modified eq, q}) one
obtains with the area of $R\left[\overline{u}\right]$ being denoted
by $A[\overline{u}]$:

\begin{eqnarray}
 & 0=-\overline{u}\left(\mathbf{r},t\right)+\int_{R\left[\overline{u}\right]}\left[w\left(\left|\mathbf{r}-\mathbf{r'}\right|\right)+q\right]d\mathbf{r'}-qA[\overline{u}_{h}]+h'=\nonumber \\
 & =-\overline{u}\left(\mathbf{r},t\right)+\int_{R\left[\overline{u}\right]}w\left(\left|\mathbf{r}-\mathbf{r'}\right|\right)d\mathbf{r'}+qA[\overline{u}]-qA[\overline{u}_{h}]+h'\label{eq: Modified, expanded}\end{eqnarray}
 implying that $\overline{u}$ is a solution of (\ref{eq: Stat states})
with the resting potential $h$ replaced by $h'-qA[\overline{u}_{h}]+qA[\overline{u}]$.
Consequently, condition 1 is satisfied. Note that the circular one-bump
solution $\overline{u}_{h}$ of the original problem (\ref{eq: Stat states})
which was used when constructing (\ref{eq: Modified eq, q}) solves
the modified problem (\ref{eq: Modified eq, q}) with $h'=h$ and
any $q$.

Eq.~(\ref{eq: Modified eq, q}) does not admit stationary solutions
with unbounded region of activation. Indeed, assuming that such a
solution $\hat{u}$ exists, substituting $\hat{u}$ into (\ref{eq: Modified eq, q})
we were to conclude that the integral term of the right-hand-side
of (\ref{eq: Modified eq, q}) is infinite which forms a contradiction.
Therefore condition 2 above is satisfied.

Finally, we show that the modification (\ref{eq: Modified eq, q})
preserves stability properties of the stationary solution $\overline{u}_{h}$
(which is a stationary solution of both the modified and the unmodified
problems (\ref{eq: Amari eq}) and (\ref{eq: Modified eq, q}) respectively
by construction). Recall that in deriving equation (\ref{eq: Eigenvals})
we did not make use of any particular assumptions on the form of the
integral kernel $w\left(\left|\mathbf{r}-\mathbf{r'}\right|\right)$.
Consequently, this expression for the eigenvalue spectrum is equally
valid for the modified problem as well as for the unmodified one.
Note, however, that when deriving stability spectrum in the case of
the modified problem (\ref{eq: Modified eq, q}) we shall exchange
$g\left(\theta-\theta'\right)$ by $g\left(\theta-\theta'\right)+qA[\overline{u}_{h}]$
(see derivations in the Appendix). Using $\int_{0}^{2\pi}qA[\overline{u}_{h}]\cos\left(n\theta\right)d\theta=qA[\overline{u}_{h}]\int_{0}^{2\pi}\cos\left(n\theta\right)d\theta=qA[\overline{u}_{h}]2\pi\delta_{n0}$
it now follows from (\ref{eq: Eigenvals}) that all eigenvalues of
the stability spectrum of $\overline{u}_{h}$ (except for $\lambda_{0}$
corresponding to perturbations of the radius of the bump) remain unaltered
by the modification. Consequently, if $\overline{u}_{h}$ is unstable
with respect to some non rotationally-invariant perturbation in the
original problem (\ref{eq: Amari eq}), it is unstable with respect
to such a perturbation in the modified problem (\ref{eq: Modified eq, q}),
whereby condition 3 holds.

The above arguments imply that a rotationally invariant solution $\overline{u}_{h}$
of Eq.~(\ref{eq: Amari eq}) that is unstable at a non-circular eigenmode
solves also the modified equation (\ref{eq: Modified eq, q}) and
is unstable with respect to the same eigenmode of the dynamics (\ref{eq: Modified eq, q}).
Furthermore, contrary to the case of (\ref{eq: Amari eq}), the dynamics
of (\ref{eq: Modified eq, q}) can never result in a periodic pattern
with unbounded activated region if $q<0$.


As stated above, conditions 1 -- 3 do not suffice to guarantee that
the activated region will remain connected as $\overline{u}_{h}$
follows the dynamics (\ref{eq: Modified eq, q}). 
Nevertheless, by tuning the parameter $q$ in (\ref{eq: Modified eq, q})
one is able to trap the dynamics in the vicinity of instability in
a state with connected activated region.

\begin{figure}[H]
 \begin{tabular}{ccc}
a\includegraphics[scale=0.19]{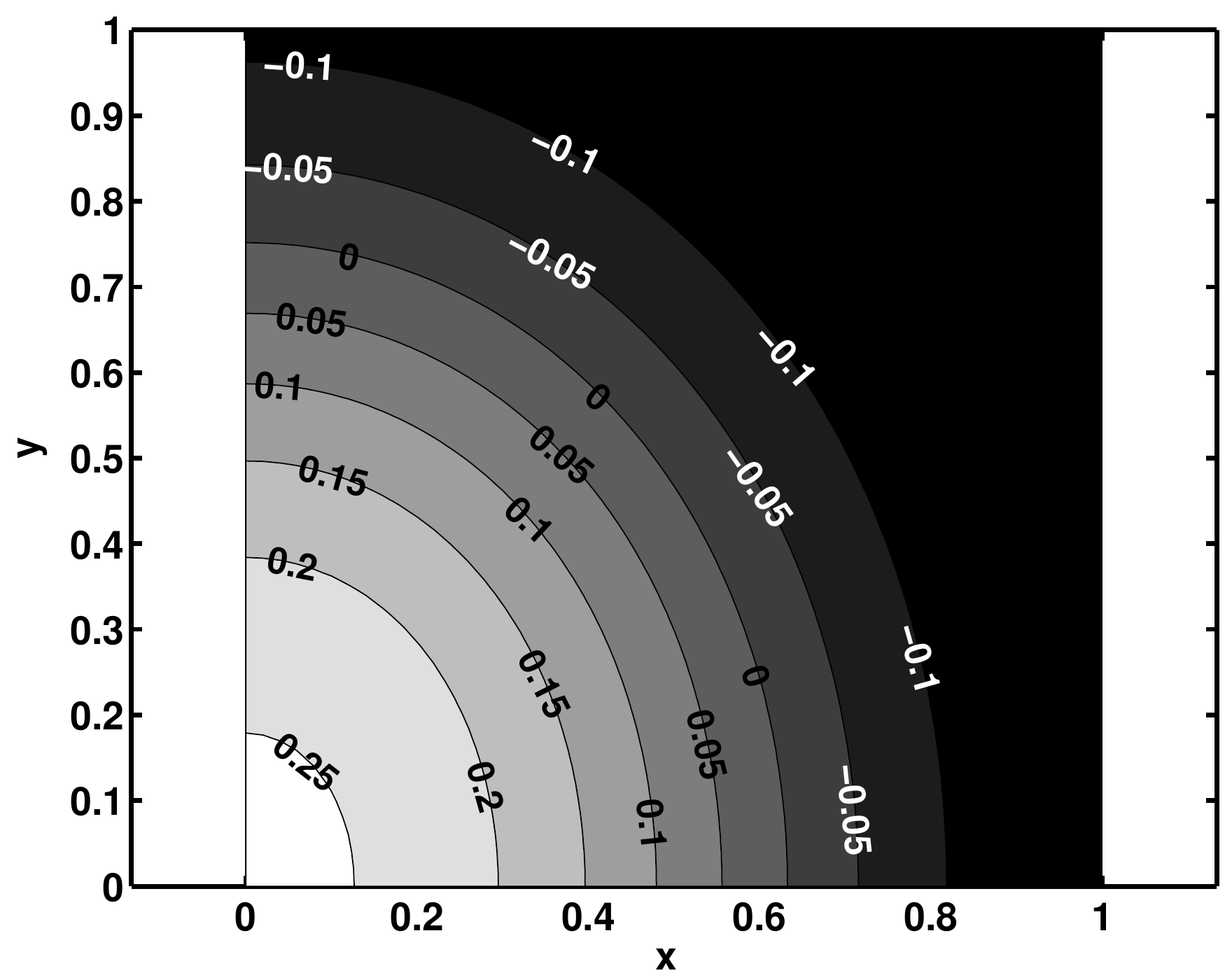}&
b\includegraphics[scale=0.19]{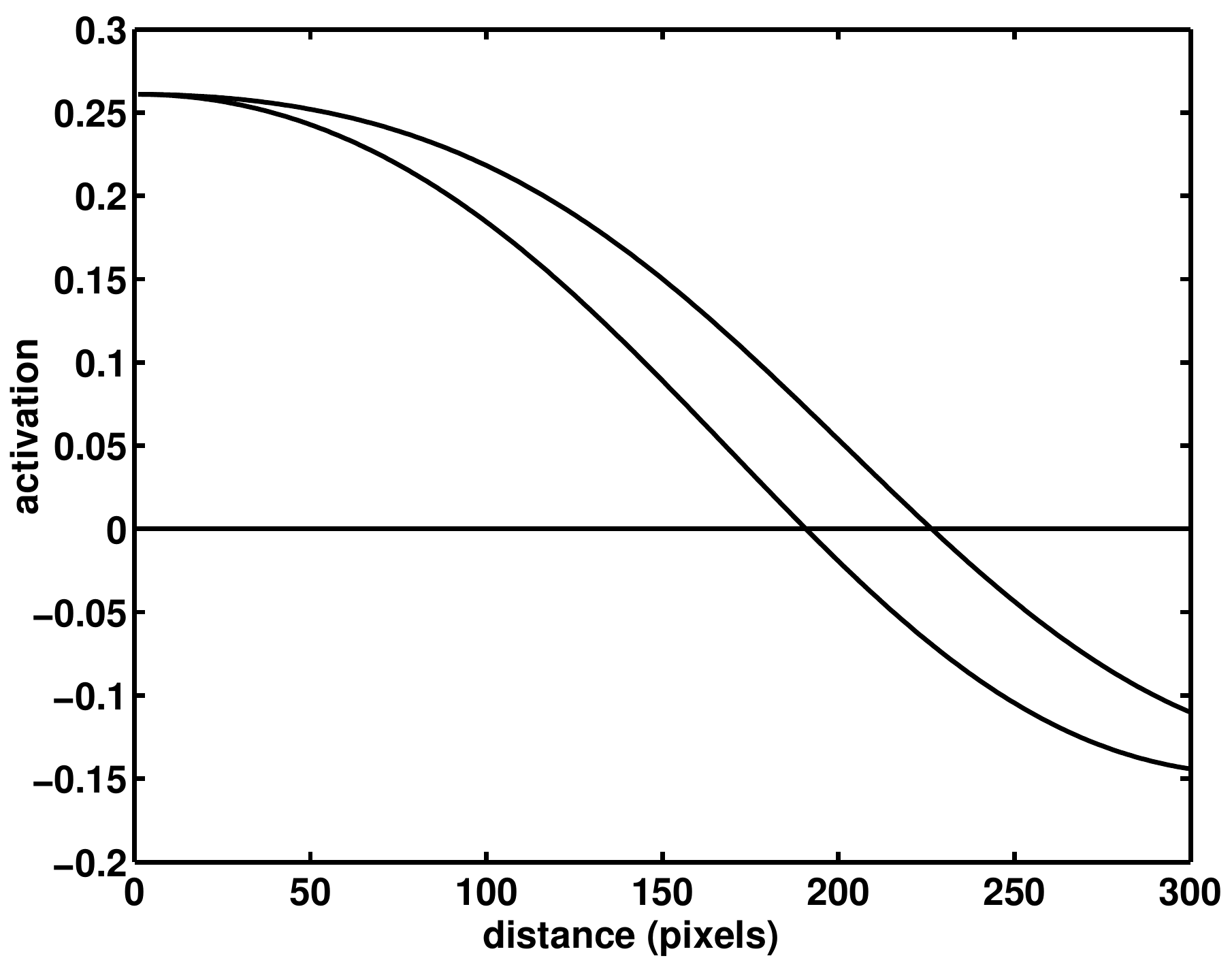}&
c\includegraphics[scale=0.19]{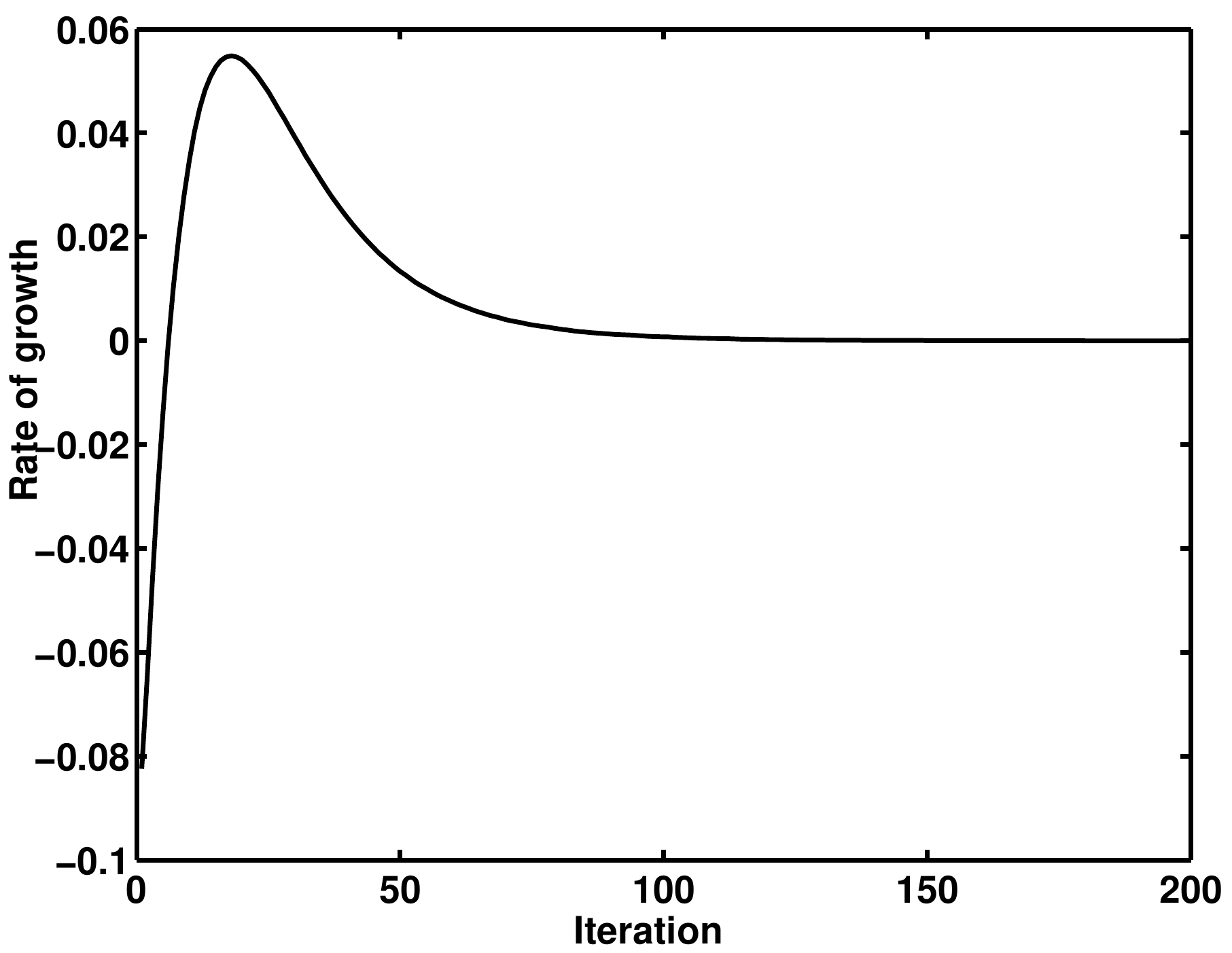}\tabularnewline
\end{tabular}

\caption{\label{fig: 6}(a) Contour 
plot of the $D_{2}$-symmetric solution superimposed with its level-curves.
The zeroes level-curve (corresponding to $u=0$) delimits the activated
region. The symmetries of the equation and initial condition allow
to restrict the simulation to one quarter of the domain. The resulting
blob is roughly ellipse-shaped with the vertical semi-axis being longer
than the horizontal one. (b) Cross-section through the profile of
the stationary solution along the coordinate axes. (c) Rate of growth
of the solution versus iteration step calculated as maximal deviation
between two subsequent time steps. Parameters are $K=1.5$, $k=5$,
$M=0.5$, $m=1.5$, $h=1.50$, $q=0.02850$.}
\end{figure}

Assume that for $q=0$ the dynamics in vicinity of the bifurcation
tends to increase the area of the activated region. Note that $q$
could be understood as a Lagrange multiplier, which ensures the constancy
of the area for one special value, which we certainly exclude. Suppose
now that $q$ is chosen to be negative. The additional term in the
integrand of (\ref{eq: Modified eq, q}) will tend to counterbalance
the area increase. We can now choose $q$ such that two effects counterbalance
and a non-rotationally invariant steady state with bounded and connected
region of activation will result. The calculation of effective values
of $q$ requires the consideration of higher orders of the dynamics
which can be expressed by a Ginzburg-Landau equation, cf.~Doubrovinski
(2005). Here, we will instead take resort to numerical simulations.
These confirm the emergence of non-rotationally invariant steady states
with bounded and connected region of activation. For example, Fig.~\ref{fig: 6}
shows the destabilization of a circular one-bump which is unstable
at a $D_{2}$-symmetric eigenmode, developing into a non-rotationally
invariant steady state with ellipse-shaped activated region. Only
one quarter of the domain was simulated (the field in the other three
quadrants is determined by that on the simulated quadrant due to Euclidean
symmetry of dynamic equations and $D_{2}$-symmetry of initial conditions)
on a grid of $300\times300$ pixels in order to increase the accuracy.
Symmetries corresponding to higher eigenvalues are irrelevant because
the eigenvalues $\lambda_{n}$ are stable for $n\ge3$ for the given
parameters. The length difference of the half-axes of the activated
domain of the resulting $D_{2}$-symmetric stationary state was much
larger than the spatial discretization step (some 30 pixels) allowing
to conclude that that the stationary solution obtained is not a discretization
artifact.

\begin{figure}[H]
\begin{tabular}{cc}
a\includegraphics[scale=0.16]{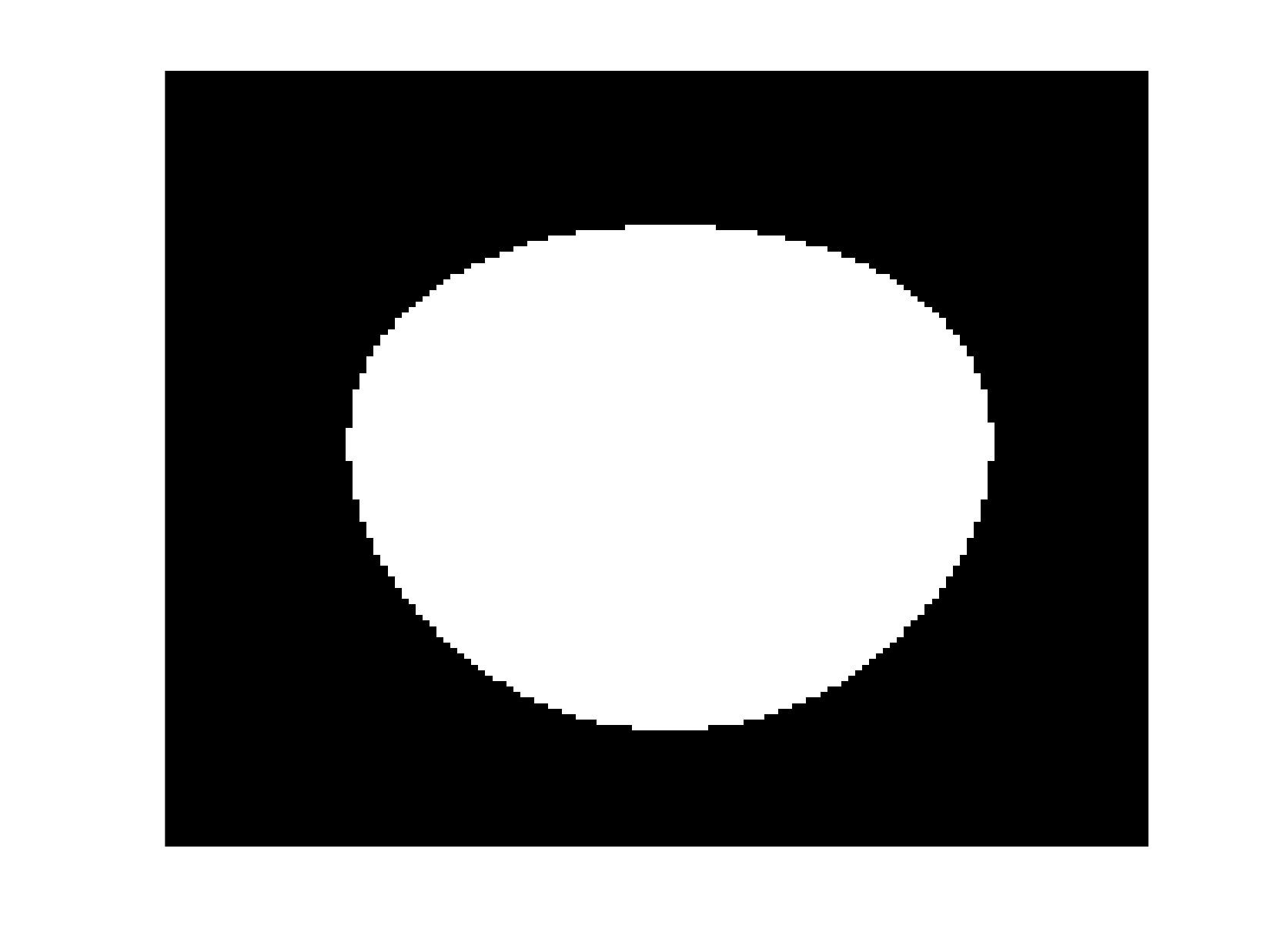}&
b\includegraphics[scale=0.16]{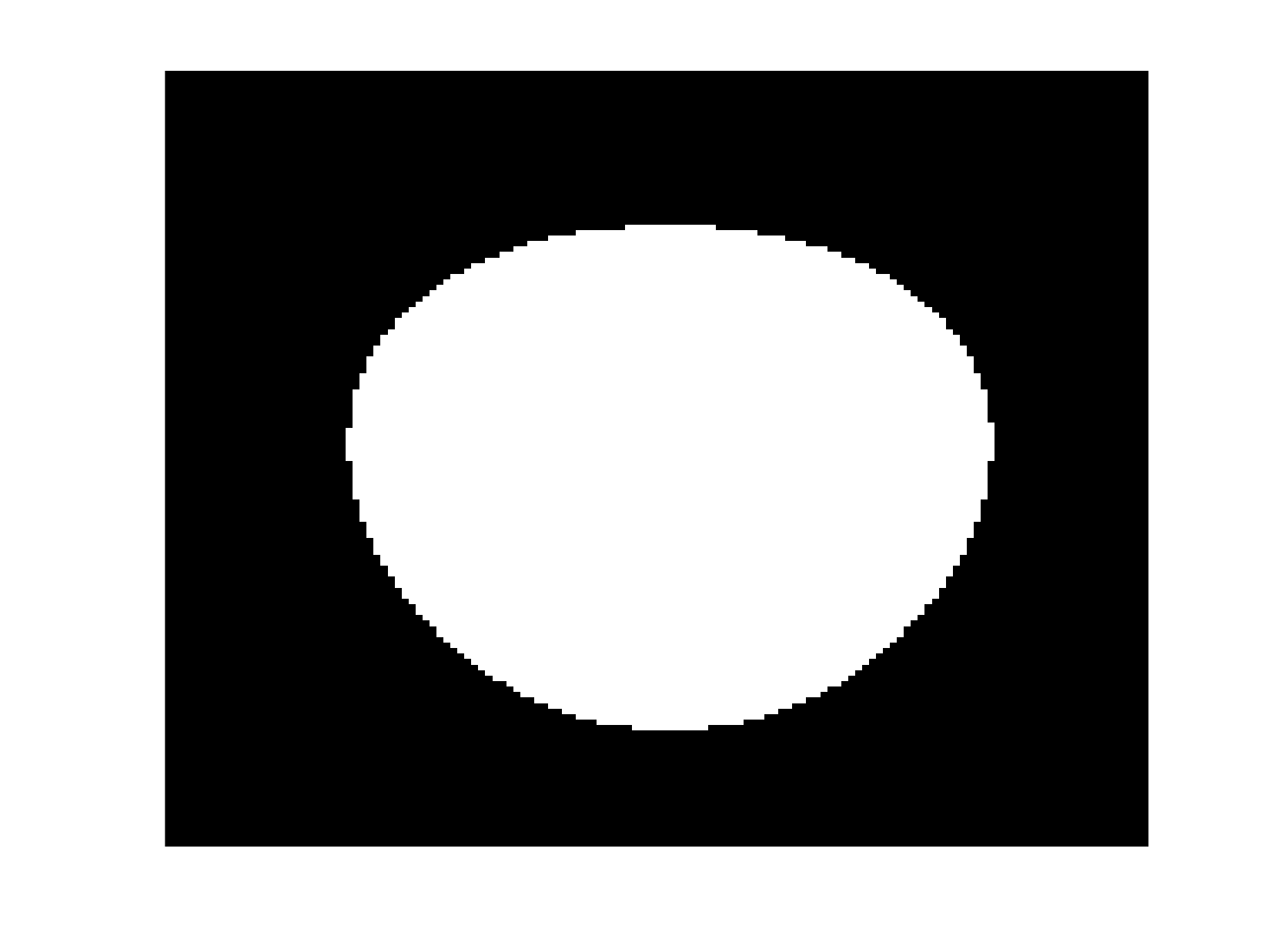}\tabularnewline
c\includegraphics[scale=0.16]{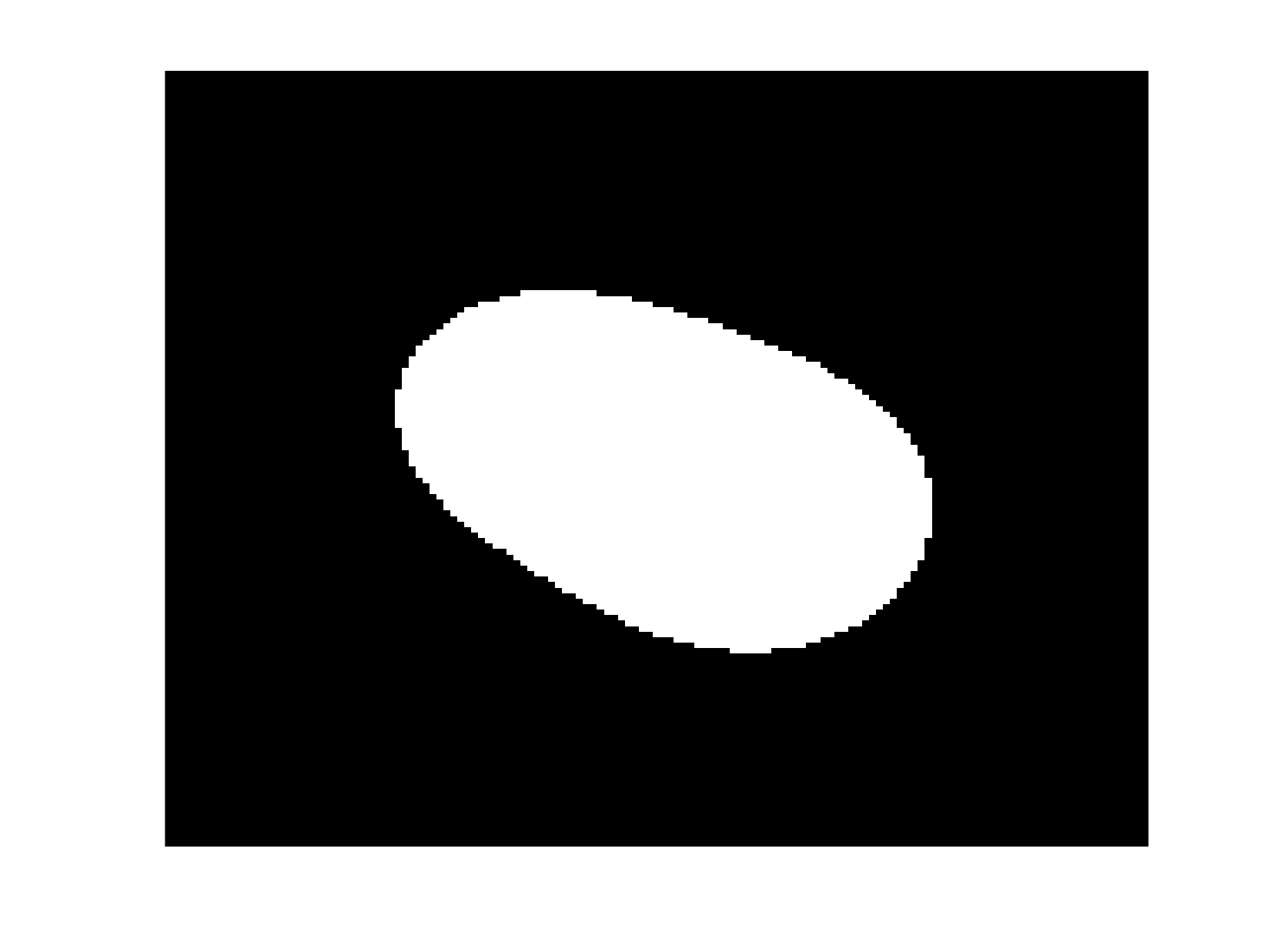}&
d\includegraphics[scale=0.16]{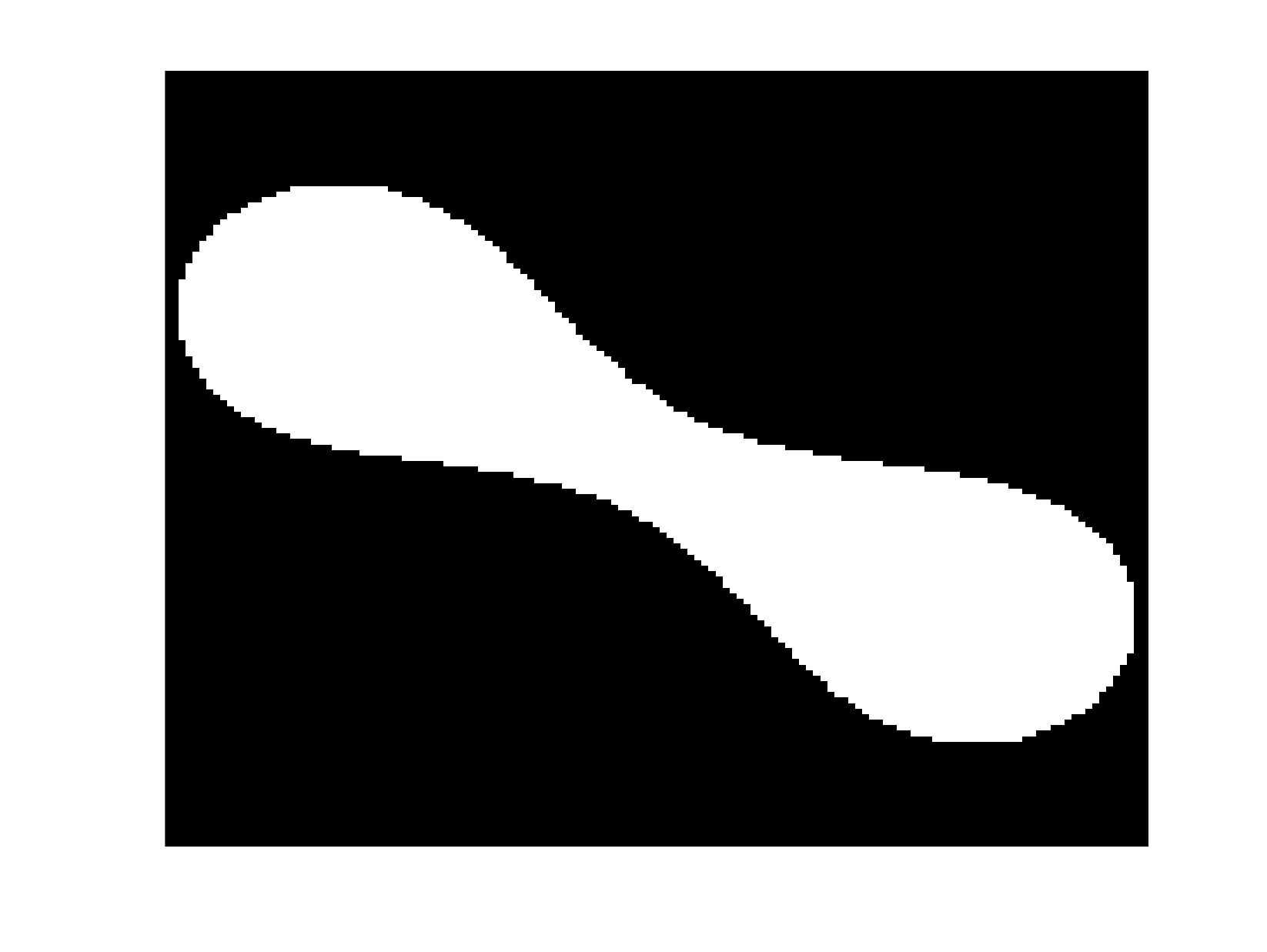}\tabularnewline
\end{tabular}\hskip 0.16cm %
\begin{minipage}[c][0pt]{0.5\textwidth}%
\begin{center}e\includegraphics[scale=0.3]{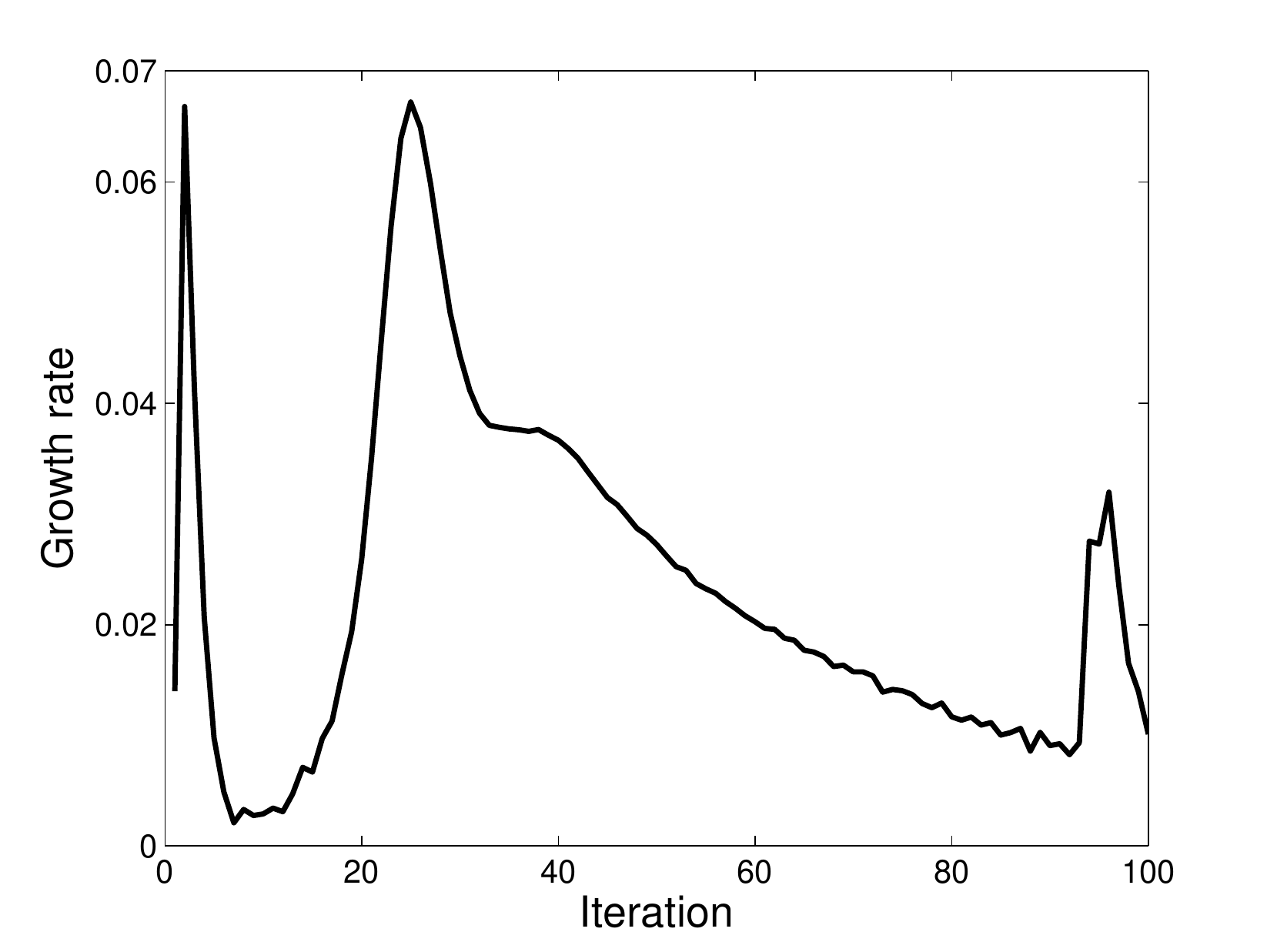} \par\end{center}%
\end{minipage}%

\caption{\label{cap: fig. 7} (a)-(d): Time evolution of activated region.
(e) Rate of growth of the solution versus iteration step calculated
as maximal deviation between two subsequent time steps. The initial
deep {}``valley'' corresponds to the shape of activated region shown
in panel (b). Parameters are: $K=1.5$, $k=5$, $M=0.5$, $m=1.5$,
$h=-0.0260$, $q=0.0125$.}
\end{figure}

Another example is given in Fig.~\ref{cap: fig. 7} showing the dynamics
in the vicinity of stability loss at a $D_{3}$-symmetric eigenmode.
Initially the field is $D_{3}$-symmetric. We see that initially a
$D_{3}$-symmetric state with connected activated region does indeed
result. The system dwells in this state during considerable time,
breaking $D_{3}$-symmetry due to small 
numerical perturbations (in absence of perturbations $D_{3}$-symmetry
must be preserved by the dynamics (\ref{eq: Modified eq, q}) due
to invariance under Euclidean transformations). Supposedly, the emerging
$D_{3}$-symmetric state is stable when the dynamical system is confined
to the subspace of $D_{3}$-symmetric functions. The final peak in
Fig.~\ref{cap: fig. 7}e, however, indicates an instability with
respect to a $D_{2}$-symmetric perturbation.

In conclusion we would like to stress that the particular choice of
modification according to (\ref{eq: Modified eq, q}) is very restrictive.
In fact, many other equations whose stationary states satisfy (\ref{eq: Amari eq})
and do not admit solutions with unbounded regions of activation are
readily constructed along the same lines. For instance $\partial_{t}u=-u+\int_{R\left[u\right]}w\left(\left|\mathbf{r}-\mathbf{r'}\right|\right)d\mathbf{r'}-\int_{R\left[u\right]}\left|\mathbf{r}-\mathbf{r'}\right|d\mathbf{r'}d\mathbf{r}+h$
can be shown to have these properties, arguing essentially as when
proving that conditions 1 and 2 above are satisfied by stationary
states of (\ref{eq: Modified eq, q}). We believe that further investigation
of such modifications will provide insights into the properties of
unstable solutions of Amari-equation.

\section{The time course of symmetry breaking}

Bifurcation theory describes the time course to critical behavior
in low-dimensional systems. Under certain conditions the parametric
destabilization of an activity distribution does not lead to a nearby
stable state, but initiates a cascade of symmetry breaking events
which eventually approaches a new distant stable configuration. It
has been a motivation for the present study to demonstrate the complex
evolution of the state of the field after the loss of stability. It
is these configurations that bear the greatest computational potentials.

For the Amari equation with a simple kernel only the existence of
rotationally invariant bumps has been proven. This work demonstrates
existence of two other non-periodic stationary states: stripe-shaped
and annular solutions. Stability analysis of these would be expected
to be very involved. Surprisingly, the special form of the Amari equation
makes this stability problem amenable to analytical treatment. The
total synaptic input to a given neuron is only dependent on the shape
of the boundary of the activated domain, making the eigenvalue problem
effectively one-dimensional, thereby allowing for fairly straight-forward
calculation of the spectrum for each of these cases.

Strictly speaking, spectral properties of the linearized operator
do not guarantee the stability of the stationary state unless additional
assumptions are satisfied. However, our extensive numerical investigation
shows that stability is indeed correctly predicted by eigenvalue analysis.
Furthermore, the spectrum allows to predict certain semiquantitative
features of solutions approached after the onset of instability.

Our numerical experiments showed that exclusively spatially extended
solutions (i.e.~those with unbounded activated domain) appeared in
the unstable parameter region. Yet, a slight modification of the interaction
kernel, introducing long-range interactions circumvents this, yielding
non-rotationally invariant stationary states with bounded and connected
region of activation which at the same time are stationary states
of the original unmodified equation. This settles a longstanding question
regarding existence of solutions of this type. This bears also relevancy
for biological systems. Assuming a certain degree of shift-twist symmetry
the existence of elongated blobs suggests mechanisms for the emergence
of orientation selectivity in neurons of the primary visual cortex.
Although the degree of asymmetry of non-rotationally invariant solutions
was moderate, these effects could in principle be enhanced by (Hebbian)
learning which we disregarded in our treatment.

\appendix

\section{Appendix}

\subsection{Circular one-bump solution}

Consider the development of a small perturbation $\varepsilon\eta$~of
a stationary rotationally-invariant one-bump~$\bar{u}$. Substituting
into the Amari equation and linearizing in~$\varepsilon$ we have
\begin{equation}
\frac{\partial\eta\left(\mathbf{x},t\right)}{\partial t}=-\eta\left(\mathbf{x},t\right)+\int_{\mathbb{R}^{2}}w\left(\left|\mathbf{x}-\mathbf{x}\mathbf{'}\right|\right)\delta\left(\bar{u}\left(\mathbf{x'}\right)\right)\eta\left(\mathbf{x'},t\right)d\mathbf{x'}.\label{eq:1a}\end{equation}
 In polar coordinates we write (somewhat informally) $w\left(\left|\mathbf{x}-\mathbf{x'}\right|\right)=w\left(r,\theta,r',\theta'\right)$,
and use the rotational invariance of~$\bar{u}$, such that \begin{equation}
\frac{\partial\eta\left(r,\theta,t\right)}{\partial t}=-\eta\left(r,\theta\right)+\int_{0}^{2\pi}\int_{0}^{\infty}r'w\left(r,\theta,r',\theta'\right)\delta\left(\bar{u}\left(r'\right)\right)\eta\left(r',\theta'\right)dr'd\theta'\label{eq:2a}\end{equation}
 Recall that\begin{equation}
\delta\left(f\left(x\right)\right)=\sum_{x_{i}}\frac{\delta\left(x-x_{i}\right)}{\left|\frac{df\left(x_{i}\right)}{dx}\right|},\label{eq:3a}\end{equation}
 where the sum is over the roots of $f$, provided that $f$ is differentiable
at the corresponding points. Using~(\ref{eq:3a}),~(\ref{eq: Stat states})
simplifies to\begin{equation}
\frac{\partial\eta\left(r,\theta\right)}{\partial t}=-\eta\left(r,t\right)+\int_{0}^{2\pi}Rw\left(r,\theta,R,\theta'\right)\frac{1}{\left|\frac{\partial\bar{u}\left(R\right)}{\partial r}\right|}\eta\left(R,\theta'\right)d\theta',\label{eq:4a}\end{equation}
 where $R$ is the radius of the (circular) region of activation of
$\bar{u}$. Substituting $\eta=e^{\lambda t}\xi\left(r,\theta\right)$
we arrive at the following eigenvalue problem. \begin{equation}
\lambda\xi\left(r,\theta\right)=-\xi\left(r,\theta\right)+\Gamma\int_{0}^{2\pi}w\left(r,\theta,R,\theta'\right)\xi\left(R,\theta'\right)d\theta',\label{eq:5a}\end{equation}
 where~$\Gamma\equiv R/\left|\left(\partial\bar{u}\left(R\right)/\partial r\right)\right|$.
By restriction of both sides to $r=R$, the eigenvalue problem~(\ref{eq:5a})
is solved by\begin{eqnarray}
 & \xi_{n}\left(R,\theta\right)=\cos n\theta\nonumber \\
 & \lambda_{n}=-1+\Gamma\int_{0}^{2\pi}w\left(R,\theta,R,0\right)\cos\left(n\theta\right)d\theta,\label{eq:6a}\end{eqnarray}
 where $n$ are nonnegative integers. The last equation can be verified
by noting that $w\left(R,\theta,R,\theta'\right)$ is a function of
$\left(\theta-\theta'\right)$ alone and Fourier-expanding the integrand
of~(\ref{eq:5a}).

The $r$-dependence of the eigenfunctions can be derived from~(\ref{eq:6a})
by exploiting the special form of the eigenvalue problem (\ref{eq:5a}).
\begin{equation}
\xi_{n}\left(r,\theta\right)=\int_{0}^{2\pi}w\left(r,\theta,R,\theta'\right)\cos\left(n\theta'\right)d\theta'\label{eq:7a}\end{equation}
 Shift invariance allows us to conclude that~$\lambda_{1}=0$. Thus,
$\Gamma$ is obtained more explicitly from Eq.~\ref{eq:6a}. \begin{equation}
\Gamma=\frac{1}{\int_{0}^{2\pi}w\left(R,\theta,R,0\right)\cos\left(\theta\right)d\theta}\label{eq:8a}\end{equation}
 Now the eigenvalues can be calculated without evaluating the slope
of the stationary solution~$\bar{u}$: \begin{equation}
\lambda_{n}=-1+\frac{\int_{0}^{2\pi}w\left(R,\theta,R,0\right)\cos\left(n\theta\right)d\theta}{\int_{0}^{2\pi}w\left(R,\theta,R,0\right)\cos\left(\theta\right)d\theta}\label{eq:9a}\end{equation}
 Eq.~(\ref{eq:9a}) is particularly convenient for examining stability
properties of circular one-bump solutions.

\subsection{Annular solutions}

The existence of solutions with annular region of activation was suggested
already in (Amari 1977). Denoting the inner radius by $R_{1}$ and
the outer radius by $R_{2}$,~(\ref{eq:4a}) is immediately rewritten
\begin{eqnarray}
\frac{\partial\eta\left(r,\theta\right)}{\partial t} & = & -\eta\left(r,\theta\right)+\Gamma_{1}\int_{0}^{2\pi}w\left(r,\theta,R_{1},\theta'\right)\eta\left(R_{1},\theta'\right)d\theta'\nonumber \\
 &  & +\Gamma_{2}\int_{0}^{2\pi}w\left(r,\theta,R_{2},\theta'\right)\eta\left(R_{2},\theta'\right)d\theta',\label{eq:10}\end{eqnarray}
 where~$\Gamma_{1}=R_{1}/\left|\partial\bar{u}\left(R_{1}\right)/\partial r\right|$,
$\Gamma_{2}=R_{2}/\left|\partial\bar{u}\left(R_{2}\right)/\partial r\right|$.
Analogously to the derivation of~(\ref{eq:5a}), we set~$\eta=e^{\lambda t}\xi\left(r,\theta\right)$
and restrict both sides to~$R_{1}$ and~$R_{2}$. \begin{eqnarray}
 & \lambda\xi_{1}=-\xi_{1}+\Gamma_{1}\int_{0}^{2\pi}w\left(R_{1},\theta,R_{1},\theta'\right)\xi_{1}'\theta'+\Gamma_{2}\int_{0}^{2\pi}w\left(R_{1},\theta,R_{2},\theta'\right)\xi_{2}'d\theta'\nonumber \\
 & \lambda\xi_{2}=-\xi_{2}+\Gamma_{1}\int_{0}^{2\pi}w\left(R_{2},\theta,R_{1},\theta'\right)\xi_{1}'d\theta'+\Gamma_{2}\int_{0}^{2\pi}w\left(R_{2},\theta,R_{2},\theta'\right)\xi_{2}'d\theta'\label{eq:11}\end{eqnarray}
 where~$\xi_{i}=\xi\left(R_{i},\theta\right)$ and $\xi_{i}'=\xi_{i}\left(\theta'\right),\, i\in\left\{ 1,2\right\} $.
It is natural to seek the eigenfunctions of the form~$\left[\xi_{1},\xi_{2}\right]=\mathbf{v}\cos\left(n\theta\right)$,
where~$\mathbf{v}$ is some two-dimensional vector. Using this substitution
it turns out that the eigenvalues of~(\ref{eq:11}) are the same
as those of the matrices\begin{equation}
\left[\begin{array}{cc}
-1+\Gamma_{1}\int_{0}^{2\pi}w_{11}\cos n\theta d\theta & \Gamma_{2}\int_{0}^{2\pi}w_{12}\cos n\theta d\theta\\
\Gamma_{1}\int_{0}^{2\pi}w_{12}\cos n\theta d\theta & -1+\Gamma_{2}\int_{0}^{2\pi}w_{22}\cos n\theta\end{array}\right]\label{eq:12}\end{equation}
 such that Eq.~\ref{eq:12} allows us to construct the spectrum of~(\ref{eq:11}).

\subsection{Stripe-shaped solutions}

Stripe-shaped solutions can be constructed by assuming the region
of activation of the form~$R\left[\bar{u}\right]=\left\{ \left(x,y\right)\mid0\leq x\leq L\right\} $
and can be seen as degenerate annuli with infinite inner radius, considered
in the previous section. Interestingly, in this case it appears possible
to give an explicit expression for the stationary one-bump solution
in terms of the model parameters:\begin{eqnarray}
 & \bar{u}\left(x,y\right)=\frac{\pi}{2km}(Km\textnormal{ erf}\left(\sqrt{k}x\right)-kM\textnormal{ erf}\left(\sqrt{m}x\right)-\nonumber \\
 & -Km\textnormal{ erf}\left(-\sqrt{k}L+\sqrt{k}x\right)+Mk\textnormal{ erf}\left(-\sqrt{m}L+\sqrt{m}x\right))-h\label{eq:13}\end{eqnarray}

The counterpart of~(\ref{eq:10}) now reads \begin{eqnarray}
\frac{\partial\eta}{\partial t} & = & -\eta+\int_{\mathbb{R}^{2}}w\left(\left|\mathbf{x}-\mathbf{x}\mathbf{'}\right|\right)\delta\left(\bar{u}\left(\mathbf{x'}\right)\right)\eta\left(d\mathbf{x'}\right)\nonumber \\
 & = & -\eta+\int_{-\infty}^{\infty}w\left(x,y,0,y'\right)\frac{1}{\left|\frac{\partial\bar{u}\left(0\right)}{\partial x}\right|}\eta\left(0,y'\right)dy'\\
 &  & +\int_{-\infty}^{\infty}w\left(x,y,L,y'\right)\frac{1}{\left|\frac{\partial\bar{u}\left(L\right)}{\partial x}\right|}\eta\left(L,y'\right)dy',\label{eq:14}\end{eqnarray}
 where~$\left(x,y\right)$ and $\left(x',y'\right)$ are Cartesian
coordinates of~$\mathbf{x}$ and~$\mathbf{x'}$ respectively. 
As in the former cases, substituting~$\eta\left(x,y,t\right)=e^{\lambda t}\xi\left(x,y\right)$
and restricting to~$0$ or to~$L$ we find \begin{eqnarray}
 & \lambda\xi_{1}=-\xi_{1}+\Gamma_{1}\int_{-\infty}^{\infty}w\left(0,0,0,y'\right)\xi_{1}'dy'+\Gamma_{2}\int_{-\infty}^{\infty}w\left(0,0,L,y'\right)\xi_{2}'dy'\nonumber \\
 & \lambda\xi_{2}=-\xi_{2}+\Gamma_{1}\int_{-\infty}^{\infty}w\left(L,0,0,y'\right)\xi_{1}'dy'+\Gamma_{2}\int_{-\infty}^{\infty}w\left(0,0,L,y'\right)\xi_{2}'dy'\label{eq:15}\end{eqnarray}
 where now $\xi_{i}'$ denotes $\xi_{i}\left(y'\right)$. Subscripts
1 and 2 designate restrictions to $x=0$ and $x=L$, respectively,
according to~$\Gamma_{1}=1/\left|\partial\bar{u}\left(0\right)/\partial x\right|$,
$\Gamma_{2}=1/\left|\partial\bar{u}\left(L\right)/\partial x\right|$
and $\xi_{1}\left(y\right)=\xi\left(0,y\right)$, $\xi_{2}\left(y\right)=\xi\left(L,y\right)$.
Arguing in exactly the same way as when considering annular solutions
we conclude that~(\ref{eq:14}) admits an uncountable infinity of
eigenvalues which are the same as those of matrices\begin{equation}
\left[\begin{array}{cc}
-1+\Gamma_{1}\hat{w}_{0}\left(\Omega\right) & \Gamma_{2}\hat{w}_{L}\left(\Omega\right)\\
\Gamma_{1}\hat{w}_{L}\left(\Omega\right) & -1+\Gamma_{2}\hat{w}_{0}\left(\Omega\right)\end{array}\right]\label{eq:16}\end{equation}
 where~$\hat{w}_{0}\left(\Omega\right)=\int_{-\infty}^{\infty}w\left(0,0,0,y'\right)\cos\left(\Omega y'\right)dy'=\int_{-\infty}^{\infty}w\left(L,0,L,y'\right)\cos\left(\Omega y'\right)dy'$,
$\hat{w}_{L}\left(\Omega\right)=\int_{-\infty}^{\infty}w\left(L,0,0,y'\right)\cos\left(\Omega y'\right)dy'=\int_{-\infty}^{\infty}w\left(0,0,L,y'\right)\cos\left(\Omega y'\right)dy'$.
To every nonnegative real number corresponds a pair of eigenvalues.
The corresponding eigenfunctions can be evaluated from~(\ref{eq:15}).

\subsubsection*{Acknowledgment}

The authors would like to thank S.~Kaijser, T.~Geisel, and S.~Amari
for their kind support of this work and for helpful remarks. We further
like to thank Hecke Schrobsdorff for stimulating discussions.

\section*{References}

\frenchspacing \raggedright \parindent -0.5cm \rule{1cm}{0cm}

S. I. Amari (1977) Dynamics of pattern formation in lateral-inhibition
type neural fields. \emph{Biological Cybernetics} \textbf{27}, 77-87.

P. Bressloff (2005) Spontaneous symmetry breaking in self-organizing
neural fields. \emph{Biological Cybernetics} \textbf{93}, 256-274.

M. Camperi, X. J. Wang (1998) A model of visuospatial working memory
in prefrontal cortex: Recurrent network and cellular bistability.
{\emph{J}ournal of Computational Neuroscience} \textbf{5}:4, 383-405.

M. G. Crandall, P. H. Rabinowitz (1971) Bifurcation from simple eigenvalues.
\textit{J. Funct. Anal.} \textbf{8}, 321-340.

D. W. Dong, J. J. Hopfield (1992) Dynamic properties of neural networks
with adapting synapses. \emph{Network: Computation in Neural Systems}
\textbf{3}:3, 267-283.

K. Doubrovinski (2005) Dynamics, stability and bifurcation phenomena
in a nonlocal model of cortical activity. http://www.matj.uu.se/research/pub/Doubrovinski1.pdf

W. Erlhagen, G. Schöner (2002) Dynamic field theory of movement preparation.
\emph{Psychological Review} \textbf{109}:3, 545-572.

W. Erlhagen, E. Bicho (2006) The dynamic neural field approach to
cognitive robotics. \emph{Journal of Neural Engineering} \textbf{3},R36-R54.

G. B. Ermentrout, J. D. Cowan (1979) A mathematical theory of visual
hallucination patterns. \emph{Biological Cybernetics} \textbf{34},
137-150.

G. B. Ermentrout (1998) Neural networks as spatiotemporal pattern
forming systems. \emph{Rep. Prog. Phys.} \textbf{61}:4, 353-430.

M. A. Giese (1998) \emph{Dynamic neural field theory for motion perception}.
Kluwer Academic Publishers, Boston.

H.-M. Gross, V. Stephan, M. Krabbes (1998) A neural field approach
to topological reinforcement learning in continuous action spaces.
\emph{IEEE World Congress on Comput. Intell.} (1998).

J. M. Herrmann, K. Pawelzik, T. Geisel (1999) Self-localization of
autonomous robots by hidden representations. \emph{Autonomous Robots}
\textbf{7}:1, 31-40.

J. M. Herrmann, H. Schrobsdorff, T. Geisel (2004) Localized activations
in a simple neural field model. \emph{CNS 2004}.

I. Iossifidis, A. Steinhage (2001) Controlling an 8 DOF manipulator
by means of neural fields. \emph{International Conference on Field
and Service Robotics} (2001).

V. K. Jirsa, K. J. Jantzen, A. Fuchs, J. A. S. Kelso (2002) Spatiotemporal
forward solution of the EEG and MEG using network modeling. \emph{IEEE
Transactions on Medical Imaging} \textbf{21}:5, 493-504.

K. Kishimoto, S. I. Amari (1979) Existence and stability of local
excitations in homogeneous neural fields. \emph{Mathematical Biology}
\textbf{7}, 303-318.

E. Kreyszig (1978) \emph{Introductory functional analysis with applications}.
Wiley, New York.

C. R. Laing, C. Chow (2001) Stationary bumps in networks of spiking
neurons. \emph{Neural Computation} \textbf{13}:7, 1473-1494.

C. R. Laing, W. C. Troy, B. S. Gutkin and G. B. Ermentrout (2002)
Multiple bumps in a neuronal model of working memory. \emph{SIAM Journal
of Applied Mathematics} \textbf{63}:1, 62-97.

C. R. Laing, W. C. Troy (2003) PDE methods for nonlocal problems.
\emph{SIAM Journal of Dynamical Systems} \textbf{2}:3, 487-516.

E. E. Lieke, R. D. Frostig, A. Arieli, D.Y . Ts'o, R. Hildesheim,
and A. Grinvald (1989) Optical imaging of cortical activity: Real-time
imaging using extrinsic dye-signals and high resolution imaging based
on slow intrinsic-signals. \emph{Annu. Rev. Physiol.} \textbf{51},
543-559.

N. Mayer, J. M. Herrmann, T. Geisel (2002) Curved feature metrics
in models of visual cortex. \emph{Neurocomputing} \textbf{44-46},
533-539.

L. M. Pismen (1999) \emph{Vortices in nonlinear fields. Form liquid
crystals to superfluids. From non-equilibrium patterns to cosmic strings.}
Clarendon Press. Oxford Science Publications.

A. Schierwagen, H. Werner (1996) Saccade control through the collicular
motor map: Two-dimensional neural field model. In: Lecture Notes In
Computer Science. Vol. 1112, Springer-Verlag, 439-444.

A. R. Schutte, J. P. Spencer, G. Schöner (2003) Testing the dynamic
field theory: Working memory for locations becomes more spatially
precise over development. \emph{Child Development} \textbf{74}:5,
1393-1417.

A. Steinhage (2000) The dynamic approach to anthropomorphic robotics.
\emph{Proceedings of the fourth Portuguese Conference on Automatic
Control, Controlo 2000}.

K. Suder, F. Wörgötter, T. Wennekers (2001) Neural field model of
receptive field restructuring in primary visual cortex. \emph{Neural
Computation} \textbf{13}, 139-159.

A. Takeuchi, S. I. Amari (1979) Formation of topographic maps and
columnar microstructures. \emph{Biological Cybernetics} \textbf{35},
63-72.

S. Tanaka, J. Ribot, K. Imamura, T. Tani (2006) Orientation-restricted
continuous visual exposure induces marked reorganization of orientation
maps in early life. \emph{NeuroImage} \textbf{30}:2, 462-477.

J. G. Taylor (1999) Neural 'bubble' dynamics in two dimensions: foundations.
\emph{Biological Cybernetics} \textbf{80}, 393-409.

E. Thelen, G. Schöner, C. Scheier, L. B. Smith (2001) The dynamics
of embodiment: A field theory of infant perseverative reaching. \emph{Behavioral
and Brain Sciences} \textbf{24}, 1-84.

H. Werner, T. Richter (2001) Circular stationary solutions in two-dimensional
neural fields. \emph{Biological Cybernetics} \textbf{85}, 211-217.

H. R. Wilson, J. D. Cowan (1973) A mathematical theory of the functional
dynamics of cortical and thalamic nervous tissue. \emph{Kybernetik}
\textbf{13}:2, 55-80.

J. Y. Wu, Y. W. Lam, C. X. Falk, L. B. Cohen, J. Fang, L. Loew, J.
Prechtl, D. Kleinfeld, Y. Tsal (1998) Voltage-sensitive dyes for monitoring
multineuronal activity in the intact central nervous system. \emph{Histochem.
J.} \textbf{30}:3, 169-187. 
\end{document}